\documentclass[iop]{emulateapj}
\def\actaa{AcA}

\begin{document}

\shorttitle{$VI$-Band Follow-Up Observations of ULPC Candidates}
\shortauthors{Ngeow et al.}

\title{$VI$-Band Follow-Up Observations of Ultra-Long Period Cepheid Candidates in M31}

\author{Chow-Choong Ngeow\altaffilmark{1}, Chien-Hsiu Lee\altaffilmark{2,3}, Michael Ting-Chang Yang\altaffilmark{1}, Chi-Sheng Lin\altaffilmark{1}, Hsiang-Yao Hsiao\altaffilmark{1}, Yu-Chi Cheng\altaffilmark{1}, Zhong-Yi Lin\altaffilmark{1}, I-Ling Lin\altaffilmark{1}, Shashi M. Kanbur\altaffilmark{4} and Wing-Huen Ip\altaffilmark{1}}
\altaffiltext{1}{Graduate Institute of Astronomy, National Central University, Jhongli 32001, Taiwan}
\altaffiltext{2}{Max Planck Institute for Extraterrestrial Physics, D-85748 Garching, Germany}
\altaffiltext{3}{University Observatory Munich, D-81679 Munich, Germany}
\altaffiltext{4}{Department of Physics, SUNY Oswego, Oswego, NY 13126, USA}

\begin{abstract}

The ultra-long period Cepheids (ULPCs) are classical Cepheids with pulsation periods exceeding $\approx 80$~days. The intrinsic brightness of ULPCs are $\sim1$ to $\sim3$ mag brighter than their shorter period counterparts. This makes them attractive in future distance scale work to derive distances beyond the limit set by the shorter period Cepheids. We have initiated a program to search for ULPCs in M31, using the single-band data taken from the Palomar Transient Factory, and identified eight possible candidates. In this work, we presented the $VI$-band follow-up observations of these eight candidates. Based on our $VI$-band light curves of these candidates and their locations in the color-magnitude diagram and the Period-Wesenheit diagram, we verify two candidates as being truly ULPCs. The six other candidates are most likely other kinds of long-period variables. With the two confirmed M31 ULPCs, we tested the applicability of ULPCs in distance scale work by deriving the distance modulus of M31. It was found to be $\mu_{M31,ULPC}=24.30\pm0.76$~mag. The large error in the derived distance modulus, together with the large intrinsic dispersion of the Period-Wesenheit (PW) relation and the small number of ULPCs in a given host galaxy, means that the question of the suitability of ULPCs as standard candles is still open. Further work is needed to enlarge the sample of calibrating ULPCs and reduce the intrinsic dispersion of the PW relation before re-considering ULPCs as suitable distance indicators.

\end{abstract}


\section{Introduction}\label{sec_intro}

Cepheid variables span a class of pulsating stars within the instability strip in the Hertzsprung-Russell diagram. Their well-known period-luminosity (PL) relation \citep[also known as the Leavitt law, as first presented in][]{leavitt1912} makes Cepheids good standard candles. They constitute the first rung of the extra-galactic distance scale ladder. Previous studies have shown that the Cepheid PL relation can be used to determine distances out to the order of 10-30~Mpc \citep[e.g.,][]{freedman2001,riess2011} using the {\it Hubble Space Telescope (HST)}. When coupled with other secondary distance indicators, such as type Ia supernovae and the Tully-Fisher relation, one is able to determine distances to galaxies that are well within the Hubble flow. However, the cosmic distance ladder suffers from several uncertainties at each ladder rung. Thus, \cite{bird2009} have proposed the use of very luminous Cepheids, the so-called ultra-long-period Cepheids (hereafter ULPCs), to extend distance estimation beyond $\approx100$~Mpc. Besides the potential to be used as distance indicators, ULPCs can also be used to constrain and enhance our understanding of stellar pulsation and evolution theories for intermediate- to high-mass stars crossing the instability strip \citep{bird2009,fiorentino2012}. \citet{bird2009} defined ULPCs to be fundamental-mode (FU) Cepheids with periods longer than $\sim80$ days. 

\begin{figure*}
  \plotone{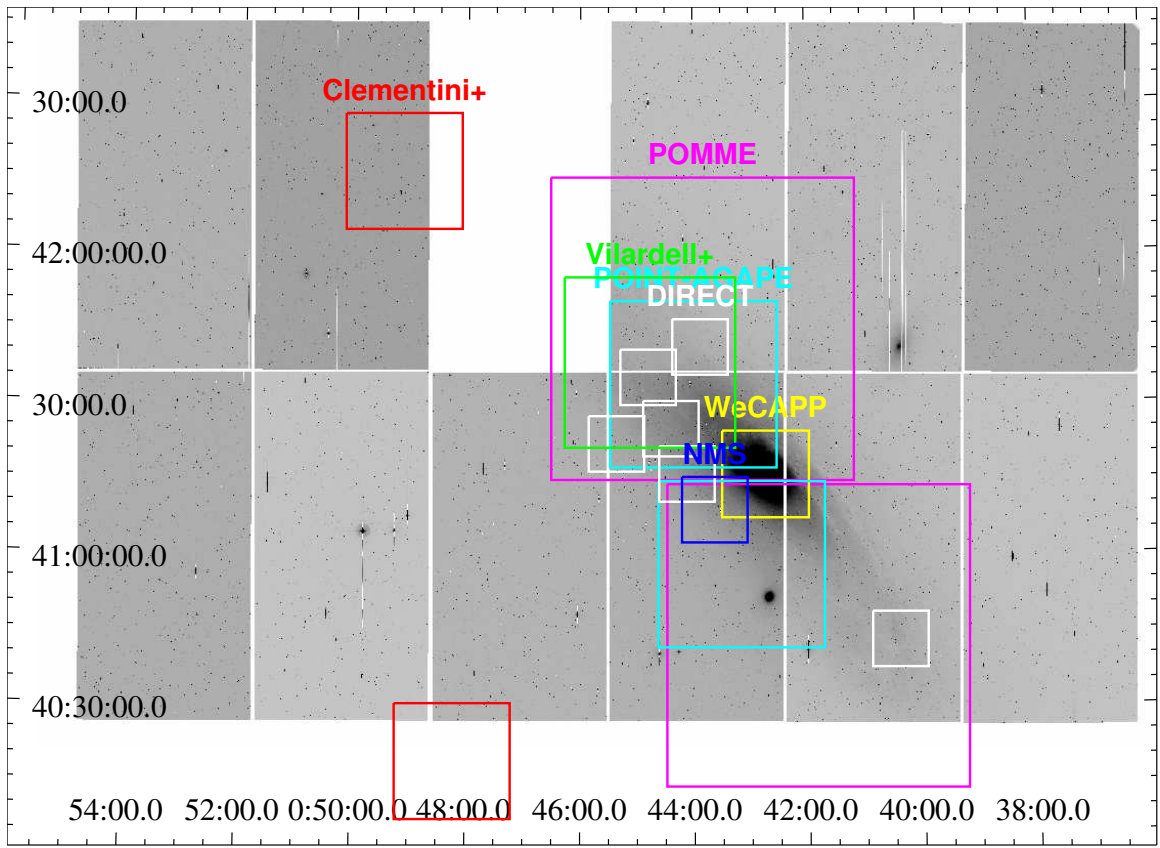}
  \caption{Footprints for some of the previous surveys, as summarized in Table~\ref{tab1}, superimposed on a PTF image of M31. Note that one CCD (upper row, third from the left) is inoperable.}
  \label{fig.surveys}
\end{figure*}

\begin{deluxetable*}{llccccl}
\tabletypesize{\scriptsize}
%\rotate
\tablecaption{Summary of Modern-day Cepheid Observations in M31. \label{tab1}}
\tablewidth{0pt}
\tablehead{
\colhead{Project/Survey\tablenotemark{a}} &
\colhead{Telescope\tablenotemark{b}} & 
\colhead{Field Center\tablenotemark{c}} &
\colhead{Field-of-View} &
\colhead{Pix. Scale\tablenotemark{d}} &
\colhead{Obs. Time Span\tablenotemark{e}} &
\colhead{Ref.\tablenotemark{f}}
}
\startdata

\cutinhead{Dedicated Surveys to Search for Cepheids (and Eclipsing Binaries)}
Freedman \& Madore& 3.6m CFHT & 10 fields covering Baade's Field I, III \& IV & $2'\times 3'$\tablenotemark{g} & $\cdots$ & 1985-1988 & 1 \\
Magnier et al.    & 2.5m INT  & $9$ fields along spiral arms & $12.5'\times 12.5'$ & 0.367 & 1993 (9) & 2 \\
                  & 1.3m MDM  & (same as above) & $10.9'\times 10.9'$ & 0.637 & 1993 (11)& 2 \\
Vilardell et al.  & 2.5m INT  & $\alpha$ = 00$^{\mathrm h}$44$^{\mathrm m}$46$^{\mathrm s}$ $\delta$ = +41$^o$38\arcmin20\arcsec & $33.8'\times 33.8'$ & 0.33  & 1999-2003 (21)     & 3, 4 \\
DIRECT: M31B      & Two Tel.\tablenotemark{h} & $\alpha$ = 11$^{\mathrm o}$.20 $\delta$ = +41$^o$.59 & $11'\times 11'$ & 0.32 & 1996-1997 ($\sim$43) & 5 \\
DIRECT: M31A      & Two Tel.\tablenotemark{h} & $\alpha$ = 11$^{\mathrm o}$.34 $\delta$ = +41$^o$.37 & $11'\times 11'$ & 0.32 & 1996-1997 ($\sim$44) & 6 \\
DIRECT: M31C      & Two Tel.\tablenotemark{h} & $\alpha$ = 11$^{\mathrm o}$.10 $\delta$ = +41$^o$.42 & $11'\times 11'$ & 0.32 & 1996-1997 ($\sim$53) & 7 \\
DIRECT: M31D      & Two Tel.\tablenotemark{h} & $\alpha$ = 11$^{\mathrm o}$.03 $\delta$ = +41$^o$.27 & $11'\times 11'$ & 0.32 & 1996-1997 ($\sim$58) & 8 \\
DIRECT: M31F      & Two Tel.\tablenotemark{h} & $\alpha$ = 10$^{\mathrm o}$.10 $\delta$ = +40$^o$.72 & $11'\times 11'$ & 0.32 & 1996-1997 ($\sim$51) & 9 \\
DIRECT: M31Y      & 1.2m FLWO & $\alpha$ = 10$^{\mathrm o}$.97 $\delta$ = +41$^o$.69 & $11'\times 11'$ & 0.33 & 1999-2000 ($\sim$25) & 10 \\
 
\cutinhead{By-products from Micro-lensing Experiments} 
AGAPE             & 2.0m TBL  & $6+1$ fields centered at M31, & $14'\times 10'$\tablenotemark{i} & 0.30 & 1994-1996 ($\sim69$) & 11 \\
                  &           & oriented along the main axis &  &  &  &  \\
NMS               & 1.0m ST   & $\alpha$ = 00$^{\mathrm h}$43$^{\mathrm m}$38$^{\mathrm s}$ $\delta$ = +41$^o$09.1\arcmin   & $13'\times 13'$     & 0.37  & 1998-2002 ($>150$) & 12, 13 \\
POINT-AGAPE       & 2.5m INT  & $\alpha$ = 00$^{\mathrm h}$43$^{\mathrm m}$10$^{\mathrm s}$ $\delta$ = +40$^o$58\arcmin15.0\arcsec   & $33'\times 33'$    & 0.33 & 1999-2001 ($\sim180$) & 14 \\
                  &           & $\alpha$ = 00$^{\mathrm h}$44$^{\mathrm m}$00$^{\mathrm s}$ $\delta$ = +41$^o$34\arcmin00.0\arcsec   & $33'\times 33'$    & 0.33 & 1999-2001 ($\sim180$) & 14 \\
WeCAPP            & 1.2m CAO  & $\alpha$ = 00$^{\mathrm h}$42$^{\mathrm m}$44$^{\mathrm s}$.3 $\delta$ = +41$^o$16\arcmin07.5\arcsec   & $17.2'\times 17.2'$ & 0.50 & 2000-2001  & 15 \\
                  & 0.8m WO   & mosaic CAO's FOV with 4 pointings & $8.3'\times 8.3'$ & 0.49 & 1999-2008 & 15 \\
POMME             & 3.6m CFHT & $\alpha$ = 00$^{\mathrm h}$43$^{\mathrm m}$50$^{\mathrm s}$ $\delta$ = +41$^o$45\arcmin0\arcsec  & $1^o\times 1^o$      & 0.187 &  2004 ($\sim50$)   & 16 \\
                  &           & $\alpha$ = 00$^{\mathrm h}$41$^{\mathrm m}$50$^{\mathrm s}$ $\delta$ = +40$^o$44\arcmin0\arcsec  & $1^o\times 1^o$      & 0.187 &  2005 ($\sim50$)   & 16 \\

\cutinhead{Other Time-series Observations}
Clementini et al. & 8.4m LBT  & $\alpha$ = 00$^{\mathrm h}$48$^{\mathrm m}$13.11$^{\mathrm s}$ $\delta$ = +40$^o$19\arcmin09.4\arcsec  & $23'\times 23'$  & 0.225 & 2007 ($\sim8$) & 17 \\
                  &           & $\alpha$ = 00$^{\mathrm h}$49$^{\mathrm m}$08.31$^{\mathrm s}$ $\delta$ = +42$^o$16\arcmin09.4\arcsec  & $23'\times 23'$  & 0.225 & 2007 ($\sim8$) & 17 \\
PAndromeda        &  1.8m PS1 & $\alpha$ = 00$^{\mathrm h}$42$^{\mathrm m}$44.33$^{\mathrm s}$ $\delta$ = +41$^o$16\arcmin07.5\arcsec  & $\sim2.6^o\times \sim2.6^o$  & 0.258 & 2010-2011 ($\sim183$) & 18 
\enddata
\tablenotetext{a}{AGAPE = Andromeda Gravitational Amplification Pixel Experiment; NMS = Nainital Microlensing Survey; POINT-AGAPE = Pixel-lensing Observations with the Isaac Newton Telescope-Andromeda Galaxy Amplified Pixels Experiment; WeCAPP = Wendelstein Calar Alto Pixellensing Project; POMME = Pixel Observations of M31 with MEgacam}
\tablenotetext{b}{CFHT = Canada-France-Hawaii Telescope (Hawaii, USA); INT = Issac Newton Telescope (Spain); MDM = McGraw-Hill Telescope at the Michigan-Dartmouth-MIT (MDM) Observatory; FLWO = F. L. Whipple Observatory (Arizona, USA); TBL = Bernard Lyot Telescope (France); WO = Wendelstein Observatory (Germany); CAO = Calar Alto Observatory (Spain); ST = Sampurnanand Telescope (India); LBT = Large Binocular Telescope (Arizona, USA); PS1 = Pan-STARRS1 Telescope (Hawaii, USA)}
\tablenotetext{c}{R.A. and decl. of field center is in $J2000$.}
\tablenotetext{d}{CCD pixel scale in ''/pixel.}
\tablenotetext{e}{Number in the parenthesis is the number of observing nights when available.} 
\tablenotetext{f}{Reference: 1. \citet{freedman1990}; 2. \citet{magnier1997}; 3. \citet{vilardell2006}; 4. \citet{vilardell2007}; 5. \citet{kaluzny1998}; 6. \citet{stanek1998}; 7. \citet{stanek1999}; 8. \citet{kaluzny1999}; 9. \citet{mochejska1999}; 10. \citet{bonanos2003}; 11. \citet{ansari2004}; 12. \citet{joshi2003}; 13. \citet{joshi2010}; 14. \citet{an2004}; 15. \citet{fliri2006}; 16. \citet{fliri2012}; 17. \citet{clementini2011}; 18. \citet{kodric2013}}
\tablenotetext{g}{The FOV for each of the 10 CCD fields.}
\tablenotetext{h}{The two telescopes used in the DIRECT project are: 1.3m MDM and 1.2m FLWO.}
\tablenotetext{i}{This is the total FOV; the FOV for individual fields is $4'\times 4.5'$.}
\end{deluxetable*}

Most studies of extragalactic distance determination use the Large Magellanic Cloud (LMC) as a distance anchor. Since the LMC has irregularity in its three-dimensional shape and low metallicity, many authors suggest instead using M31 as a stepping stone to cosmic-distance determination \citep[see, e.g.,][and references therein]{clementini2001,vilardell2010}. The merits of M31 include the following: it has a simple geometry; potential M31 distance-indicator stars are bright enough to be resolved; it is a local counterpart to the spiral galaxies that are used to determine the extragalactic distance \citep[see, e.g.,][]{freedman2001}; and it is a local benchmark to calibrate the Tully-Fisher relation. Despite its proximity, M31 has not been considered as a distance anchor \citep[see, e.g.,][]{riess2011}, because a distance anchor has to have a precise distance estimate and harbor a fair amount of distance indicators such as Cepheids. M31 has not yet met these two criteria because (1) its current distance estimate has larger uncertainty \citep[$\sim4\%$, ][]{vilardell2010} compared to other distance anchors \citep[for example, LMC at $\sim2\%$ level; see][]{pietrzynski2013}, and (2) there was no large sample of well studied Cepheids in the literature. Nevertheless, recent studies of M31 eclipsing binaries \citep{lee2014b} and Cepheids \citep{fliri2012,riess2012,kodric2013,kodric2014} have demonstrated the potential of establishing M31 as a distance anchor in the near future. Given the increasing importance of M31 in future distance scale work, it makes sense to search for and identify ULPCs in M31: such studies could provide a ``one-step'' calibration of the Hubble constant, similar to the role of NGC 4258 in \citet{riess2011}.

Ground-based CCD observations of Cepheids in M31 originate either from dedicated surveys to search for Cepheids and (detached) eclipsing binaries, or as by-products from intense monitoring of M31 to detect micro-lensing events. Examples of the former case include \citet{freedman1990}, \citet{magnier1997}, and \citet{vilardell2006,vilardell2007}, as well as the DIRECT project \citep[see][and subsequent papers in the series]{kaluzny1998}. For the micro-lensing experiments, these include the AGAPE \citep{ansari2004}, the Nainital Microlensing Survey \citep{joshi2010}, the POINT-AGAPE Survey \citep{an2004}, the WeCAPP survey \citep{fliri2006}, and the POMME Survey \citep{fliri2012}. In addition, time-series observations of two M31 fields, using the Large Binocular Telescope, have also detected a number of short-period Cepheids \citep{clementini2011}. Recently, \citet{riess2012} combined data from the POMME Survey and {\it HST} observations for 68 Cepheids, and derive a true distance of $752\pm27$~kpc to the M31. \citet{kodric2013} presented a catalog of M31 Cepheids, including fundamental and first overtone Cepheids (as well as Type II Cepheids), based on the first year of the Panoramic Survey Telescope and Rapid Response System (PS1, Pan-STARRS1) PAndromeda Survey. A summary of these surveys and projects is presented in Table~\ref{tab1}, and the coverage for some of the surveys is shown in Figure~\ref{fig.surveys}. These studies are limited in that most of the observations, with the exception of the POMME Survey and the PAndromeda Survey, cover only part of M31, and most of them concentrate on the disk (see Figure~\ref{fig.surveys}). Furthermore, no Cepheids with periods longer than $\sim80$ days were detected in these studies.

\begin{figure*}
  \plotone{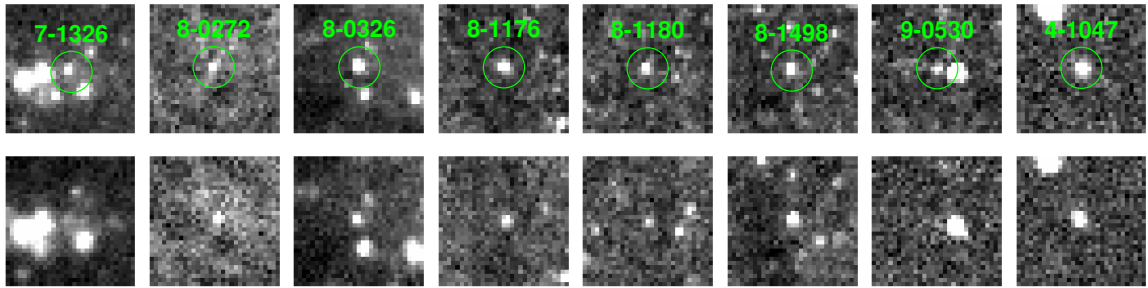}
  \caption{Postage-stamp images of the eight ULPC candidates reported in Paper I. The upper row shows the PTF $R$-band images, while the lower row shows the PTF $g$-band images. Note that the $R$- and $g$-band images are from different epochs. All stamps have a FOV of $30'' \times 30''$. The green circles with a radius of $5''$ indicate the locations of the candidates.}
  \label{fig.stamp}
\end{figure*}

Therefore, we have initiated a program to search for ULPCs in M31 by using data from the Palomar Transient Factory \citep[PTF;][]{law2009,rau2009}. This is because the PTF data can cover the entire disk of M31, as shown in Figure~\ref{fig.surveys}. In addition, PTF observes M31 routinely with cadence up to one day. These two conditions make PTF an ideal data set to search for the ULPCs in M31. Our search results were published in \citet[][hereafter Paper I]{lee2013} and \citet{lee2014}. Using image subtraction techniques, eight ULPC candidates were identified. Figure \ref{fig.stamp} displays the postage stamp images for these candidates based on PTF data. As mentioned in Paper I, time series $VI$-band data is needed to further confirm or disprove the ULPC nature of these candidates. In this work, we report the $VI$-band follow-up observations for all these eight candidates. Section 2 describes the follow-up observations, data reduction, and the calibration of the light curves. Analysis of these light curves and the results are presented in Section 3, followed by our conclusions in Section 4.

\section{The Follow-Up Observations}\label{sec_obs}

\begin{figure}
  \plotone{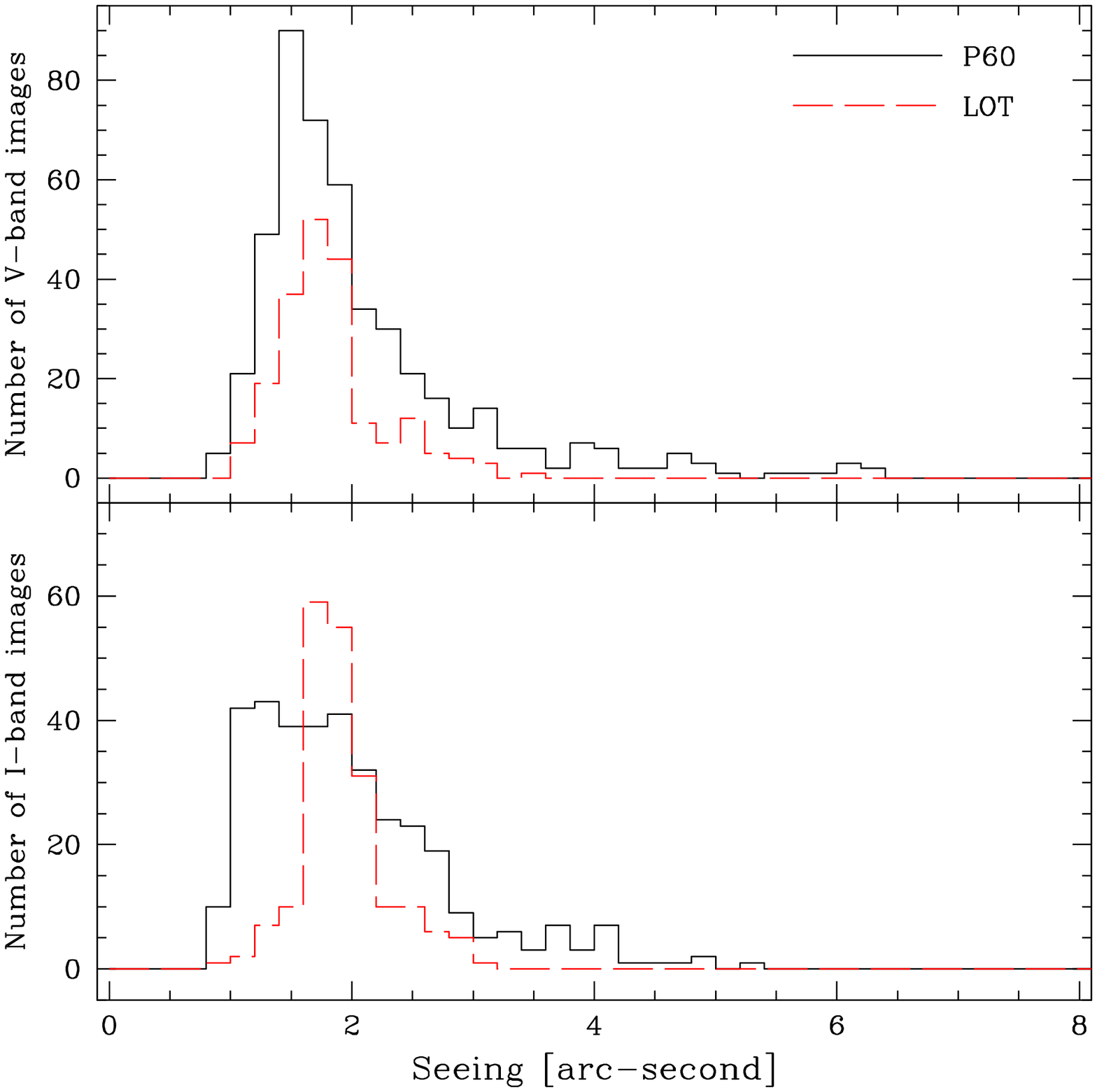}
  \caption{Seeing distributions in the $V$ (upper panel) and $I$ bands (lower panel) for the P60 and LOT images. For the P60 observations, besides a pair of $VI$-band images, an additional $V$-band image was also taken on the nights between 2012 November and 2013 March.}
  \label{fig.seeing}
\end{figure}

The $VI$-band follow-up observations were carried out with the P60 Telescope \citep[P60;][]{cenko2006} at the Palomar Observatory and the Lulin 1 m Telescope \citep[LOT;][]{kinshita2005} at Lulin Observatory. These observations began in 2012 October and ended in 2013 November. P60 is equipped with a $2\mathrm{k}\times 2\mathrm{k}$ CCD with a pixel scale of $0.379$ arcsec per pixel, while LOT used a $1\mathrm{k}\times 1\mathrm{k}$ CCD with a pixel scale of $0.512$ arcsec per pixel. Observations for both telescopes were obtained in queue mode when there was available observing time in suitable weather conditions. Imaging data from P60 was reduced with a dedicated pipeline as described in \citet{cenko2006}. For LOT images, subroutines in IRAF\footnote{IRAF is distributed by the National Optical Astronomy Observatories, which are operated by the Association of Universities for Research in Astronomy, Inc., under cooperative agreement with the National Science Foundation.} were used to reduce the data in the usual manner (bias and dark subtraction, and flat-fielded with master flat images). The LOT $I$-band images suffered from a fringing problem. To remove the fringe patterns, we median combined the $I$-band images for candidates with an internal ID 4-1047\footnote{See Paper I for the meaning of internal ID.} and created a master fringe image. We selected the images for this candidate because it is located far away from the M31 bulge and disk (see Figure 1 in Paper I). Hence, the images do not have a strong background gradient. The master fringe image was smoothed by a ``boxcar'' averaging algorithm available from IRAF, and then scaled and subtracted from all of the LOT $I$-band images. Astrometric refinement for both P60- and LOT-reduced images were performed using {\tt astrometry.net} \citep{lang2010}. Figure~\ref{fig.seeing} shows the seeing distributions of all the images taken from both telescopes. It can be seen that some images with large seeing could be affected by bad weather. After visual inspection, images that were affected by bad weather were discarded. This left 568 and 386 images from P60 and LOT, respectively.

Photometry based on point-spread function fitting (hereafter PSF photometry) for the candidates and suitable stars in each frame were obtained from {\tt IRAF/DAOPHOT} subroutines. For each candidate, we first created a ``master'' catalog (with $\sim50$ stars) that included: (1) the location of the ULPC candidate itself; (2) faint stars in the vicinity of the candidate star (selected based on the best seeing images); and (3) relatively bright ($V_{LGS}\sim16.3\mathrm{mag.}$ to $\sim19.9\mathrm{mag.}$) and isolated stars taken from the M31 Local Group Survey catalog \citep[LGS;][]{massey2006}. These LGS stars were visually inspected to ensure they were located within the P60 and LOT images, as well as away from any crowded regions in the images. About $\sim15$ to $\sim25$ LGS stars in the ``master'' catalog were used to construct the PSF model for each image by executing the IRAF subroutines {\tt DAOPHOT.PHOT}, {\tt DAOPHOT.PSTSELECT} and {\tt DAOPHOT.PSF}. Sky coordinates for all the stars in the ``master'' catalogs were then converted to pixel coordinates and saved as a coordinate file. Finally, {\tt DAOPHOT.PHOT} and {\tt DAOPHOT.ALLSTAR} were run with the input coordinate file and PSF model to obtain the instrumental PSF photometry for the stars in the ``master'' catalog. 

\subsection{Light Curves from Differential Photometry}

The $VI$-band light curves for the eight candidates were constructed using differential photometry techniques given in \citet{broeg2005}. Instead of using a single star as a comparison star (cs) to obtain differential photometry, the technique presented in \citet{broeg2005} used a number of stars (either all stars in the images or a subset of suitable constant stars) to construct an artificial comparison star:

\begin{eqnarray}
\Delta m & = & m_c -<m>_{\mathrm{cs}} \ = \ m_c - \sum_i w_i m_i^{\mathrm{cs}},
\end{eqnarray}

\noindent where $m_c$ was the instrumental magnitudes for our candidates in either the $V$ or $I$ band, and $w_i$ were the weights for individual comparison stars such that $\sum_i w_i=1$. At the first stage, we constructed differential photometry light curves for a pair of LGS stars in our ``master'' catalogs, as mentioned in the previous section, by taking one of them as a comparison star. We then removed those LGS stars that exhibited a large scatter in the differential light curves. This left a subset of good LGS stars before equation (1) was applied to each of our candidates. Since the calibrated $VI$-band magnitudes were available for these comparison stars from the LGS catalog, light curves constructed from equation (1) were calibrated via the following equation:

\begin{eqnarray}
m_{\mathrm{calibrate}} & = & \Delta m + \sum_i w_i m_i^{\mathrm{cs,LGS}}, 
\end{eqnarray}

\begin{figure*}
  \plottwo{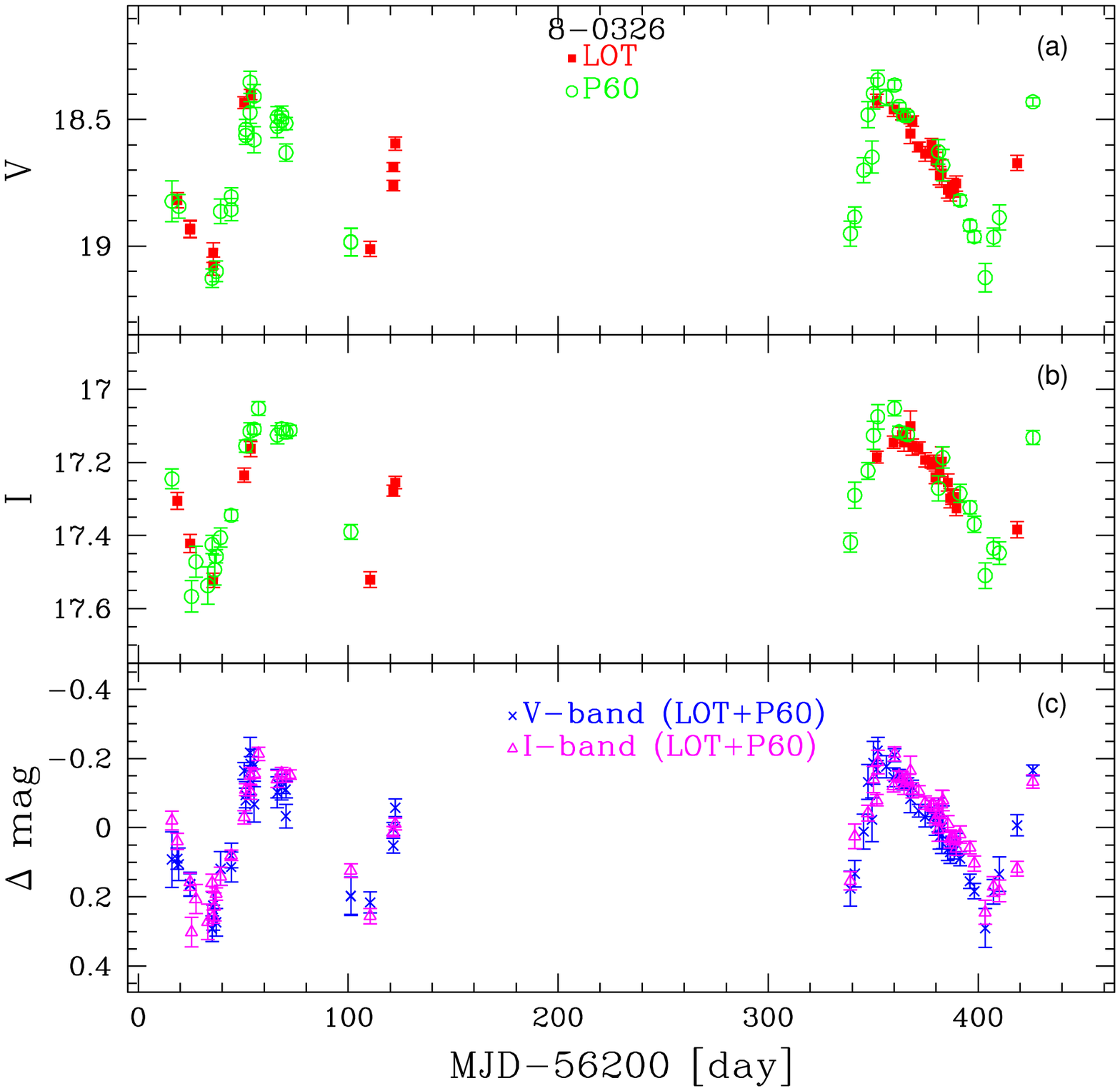}{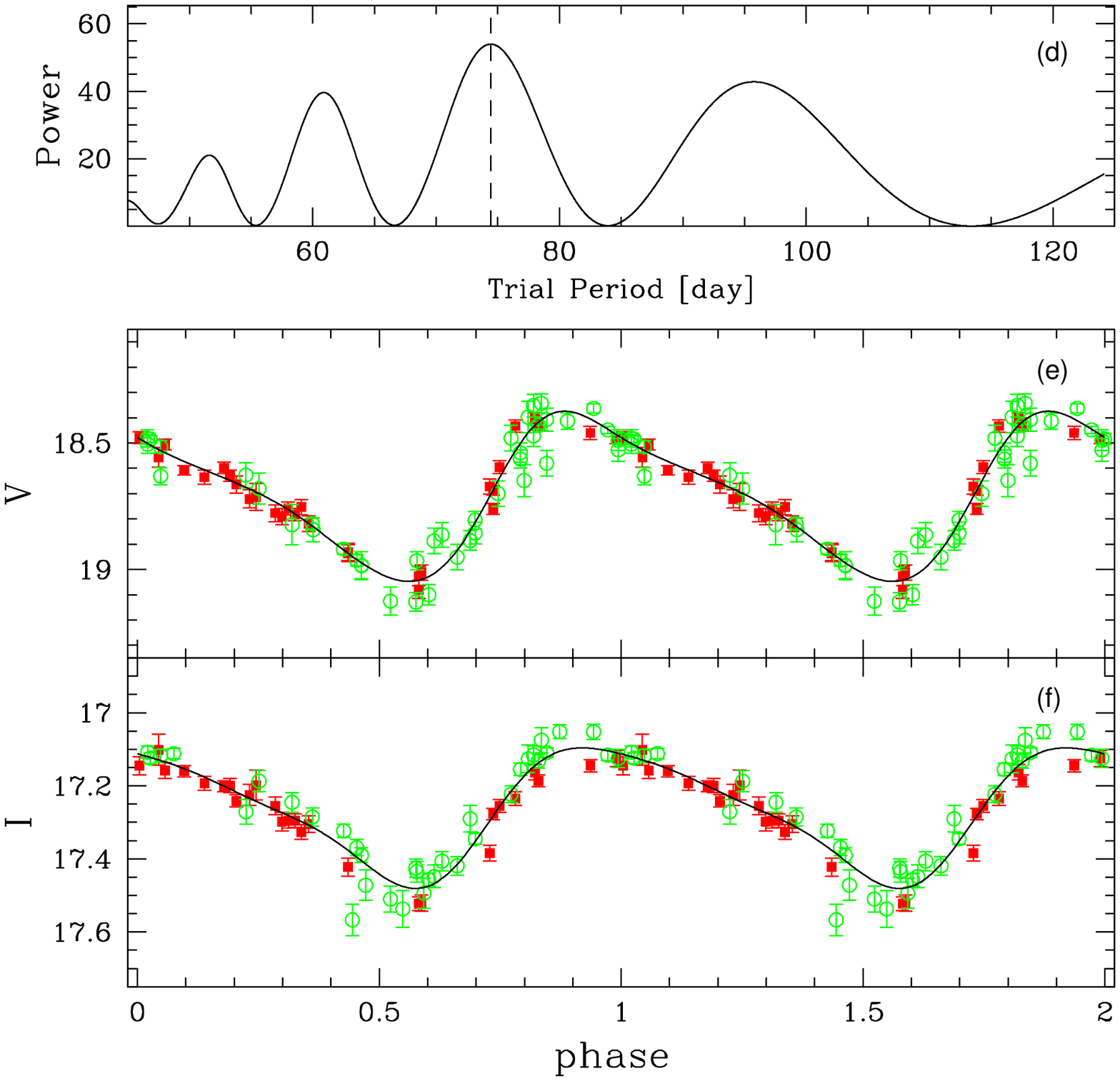}
  \caption{{\bf (a)} Observed $V$-band light curve for candidate 8-0326, where the green open circles and red filled squares are for the P60 and LOT data, respectively. {\bf (b)} Observed $I$-band light curve for the same candidate. {\bf (c)} the combined $VI$ band light curve that was used in period search; see the text (Section 3) for details. {\bf (d)} The Lomb-Scargle periodogram, at which the corresponding period at the peak of the periodogram (indicates by a vertical dashed line), is the adopted period listed in Table \ref{tab_result}. {\bf (e)} The $V$-band folded light curve using the period given in Table \ref{tab_result}. {\bf (f)} The $I$-band folded light curve for the same candidate. The black curves in (e) and (f) represent the fitted light curves based on low-order Fourier decomposition.} \label{fig_80326}
\end{figure*}

\begin{figure*}
  \plottwo{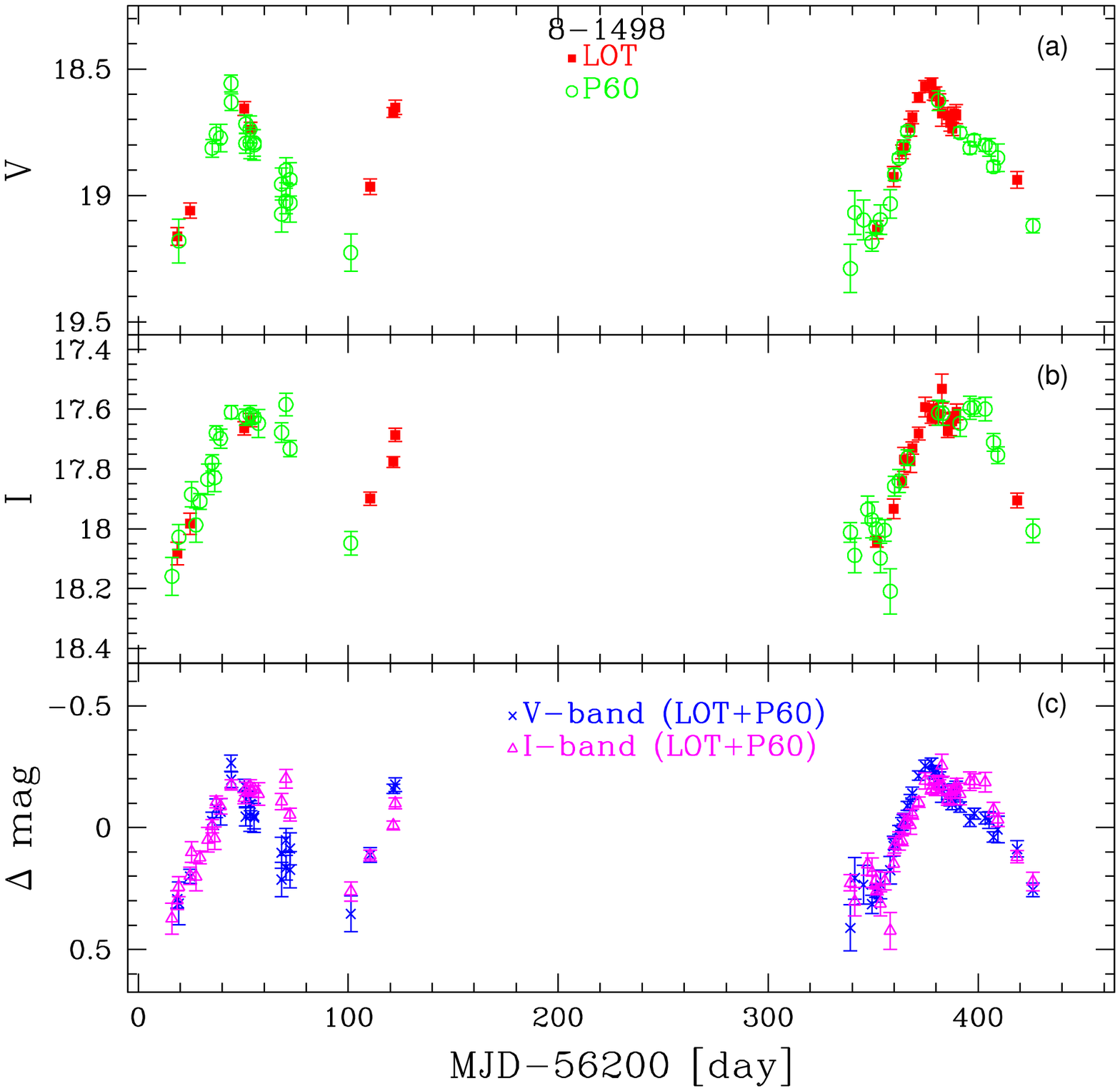}{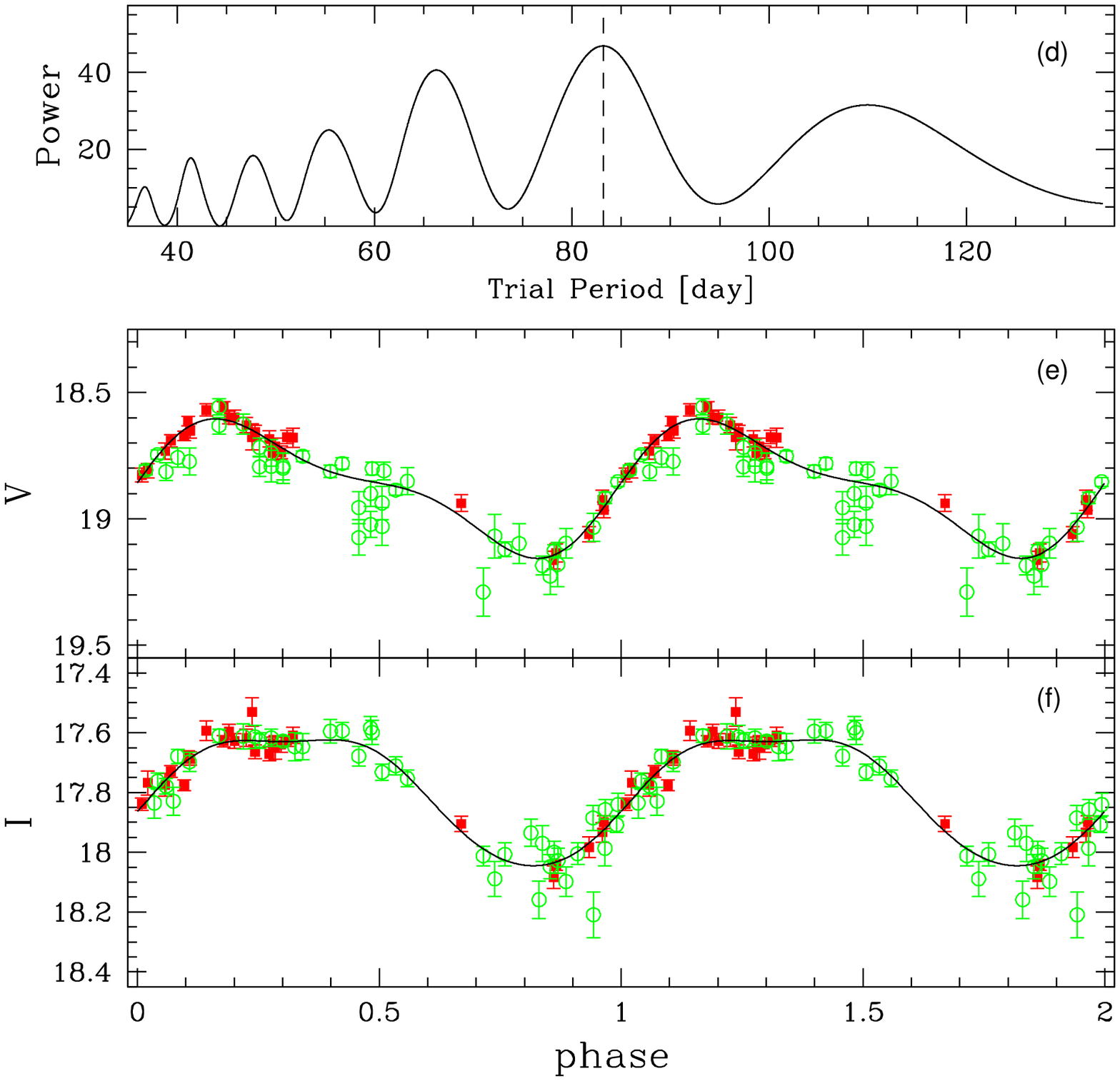}
  \caption{Same as Figure \ref{fig_80326}, but for candidate 8-1498.} \label{fig_81498}
\end{figure*}

\begin{figure*}
  $\begin{array}{cc}
    \includegraphics[angle=0,scale=0.40]{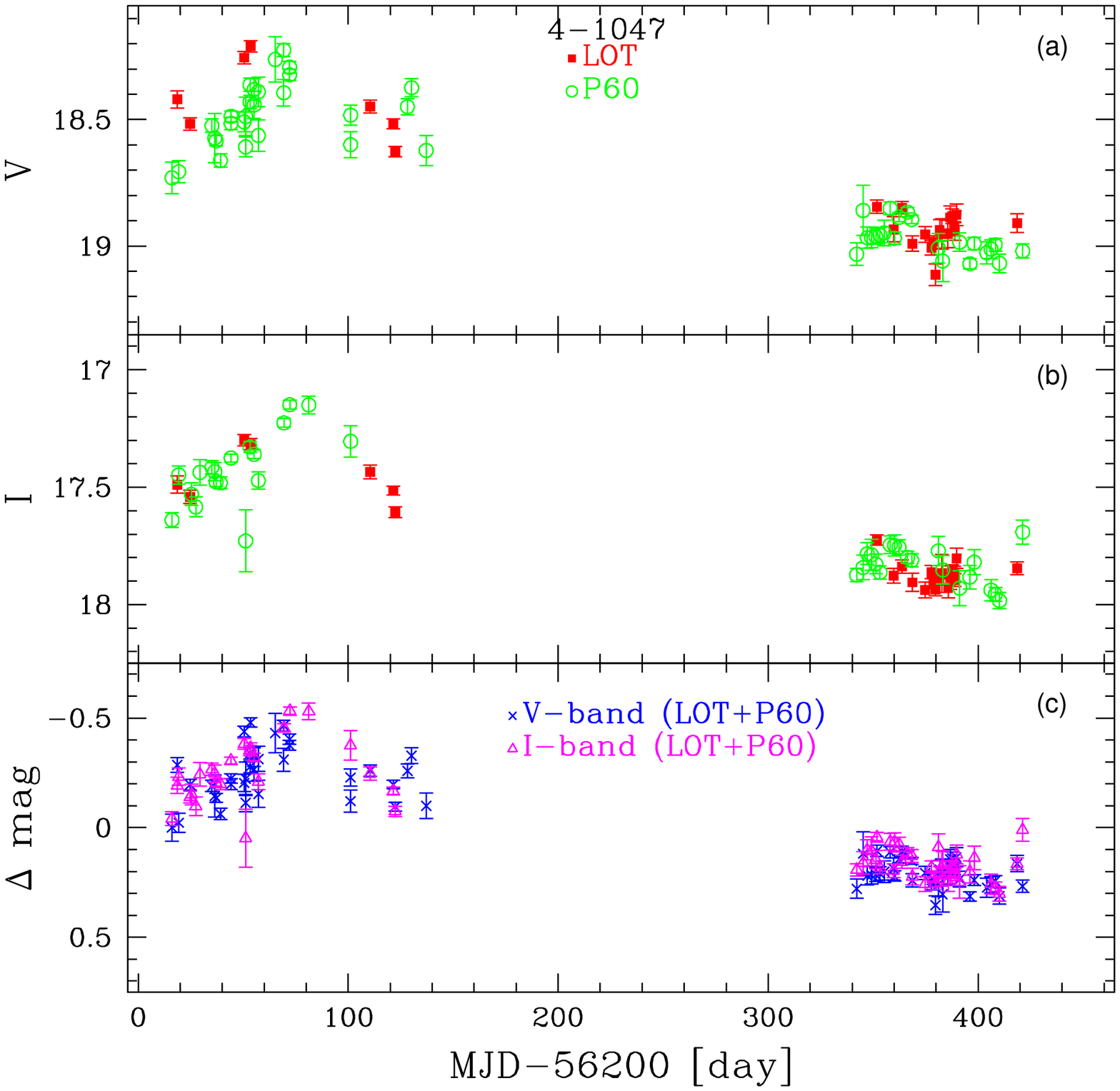} & 
    \includegraphics[angle=0,scale=0.40]{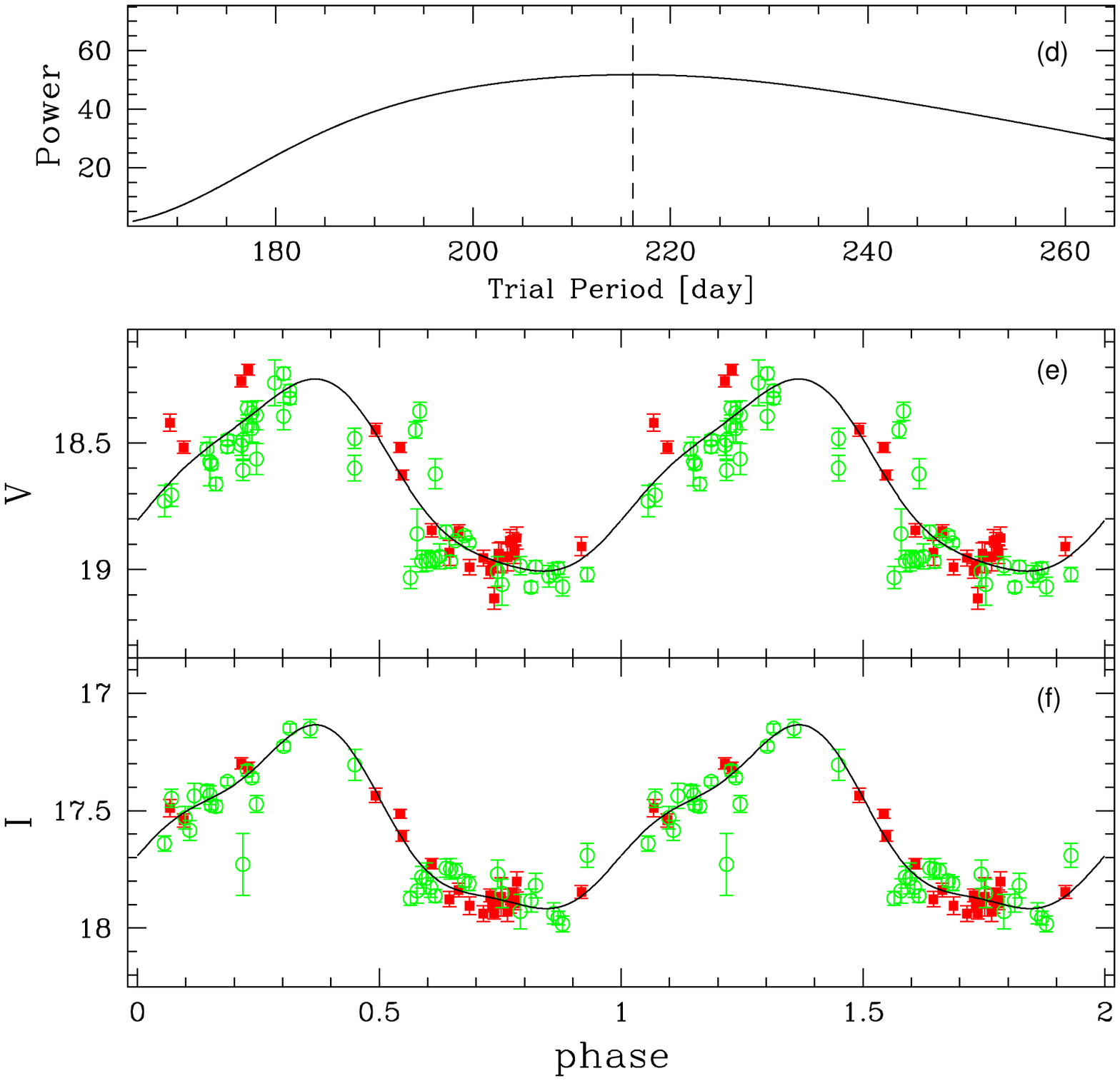} \\
    \includegraphics[angle=0,scale=0.40]{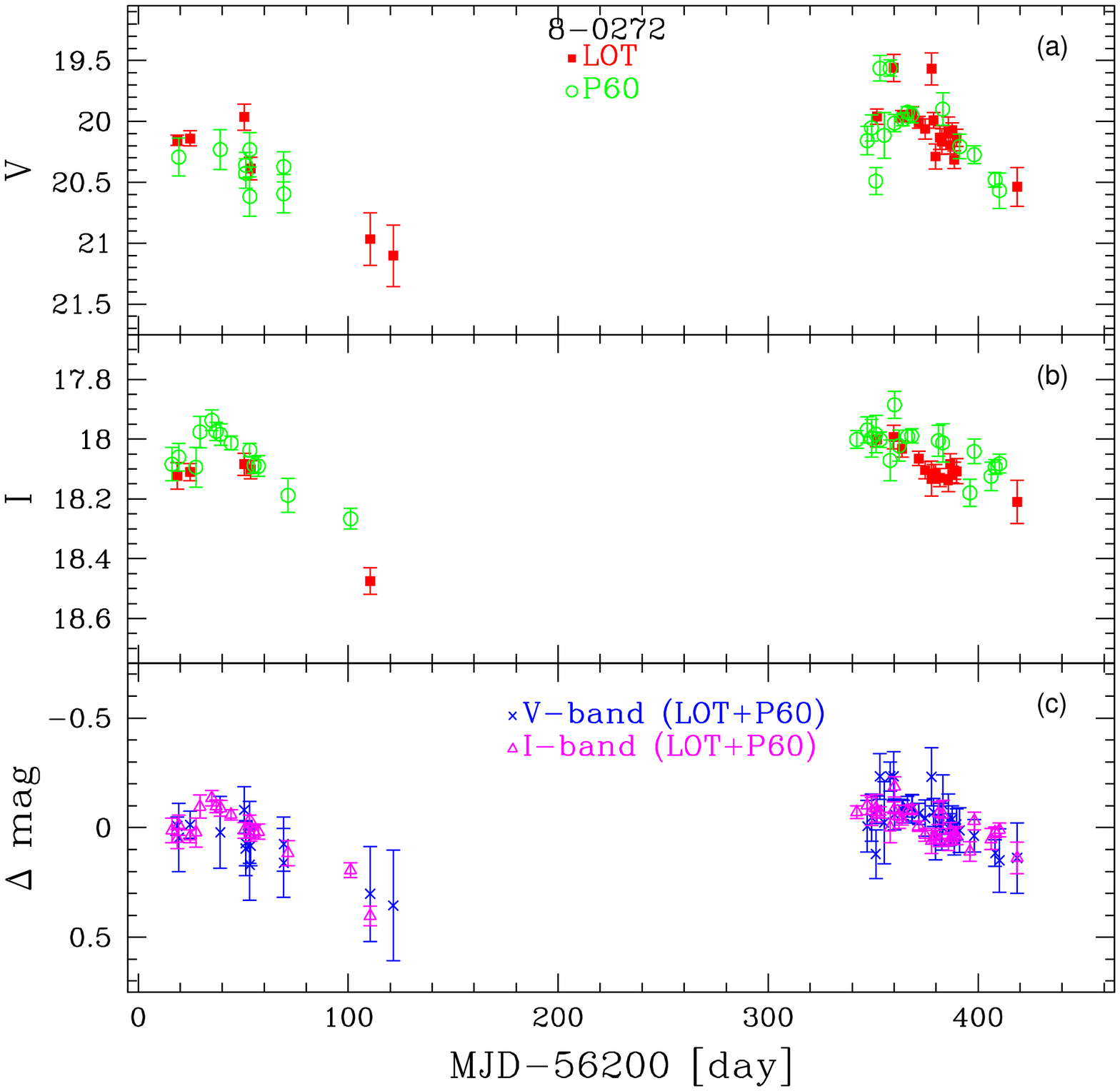} & 
    \includegraphics[angle=0,scale=0.40]{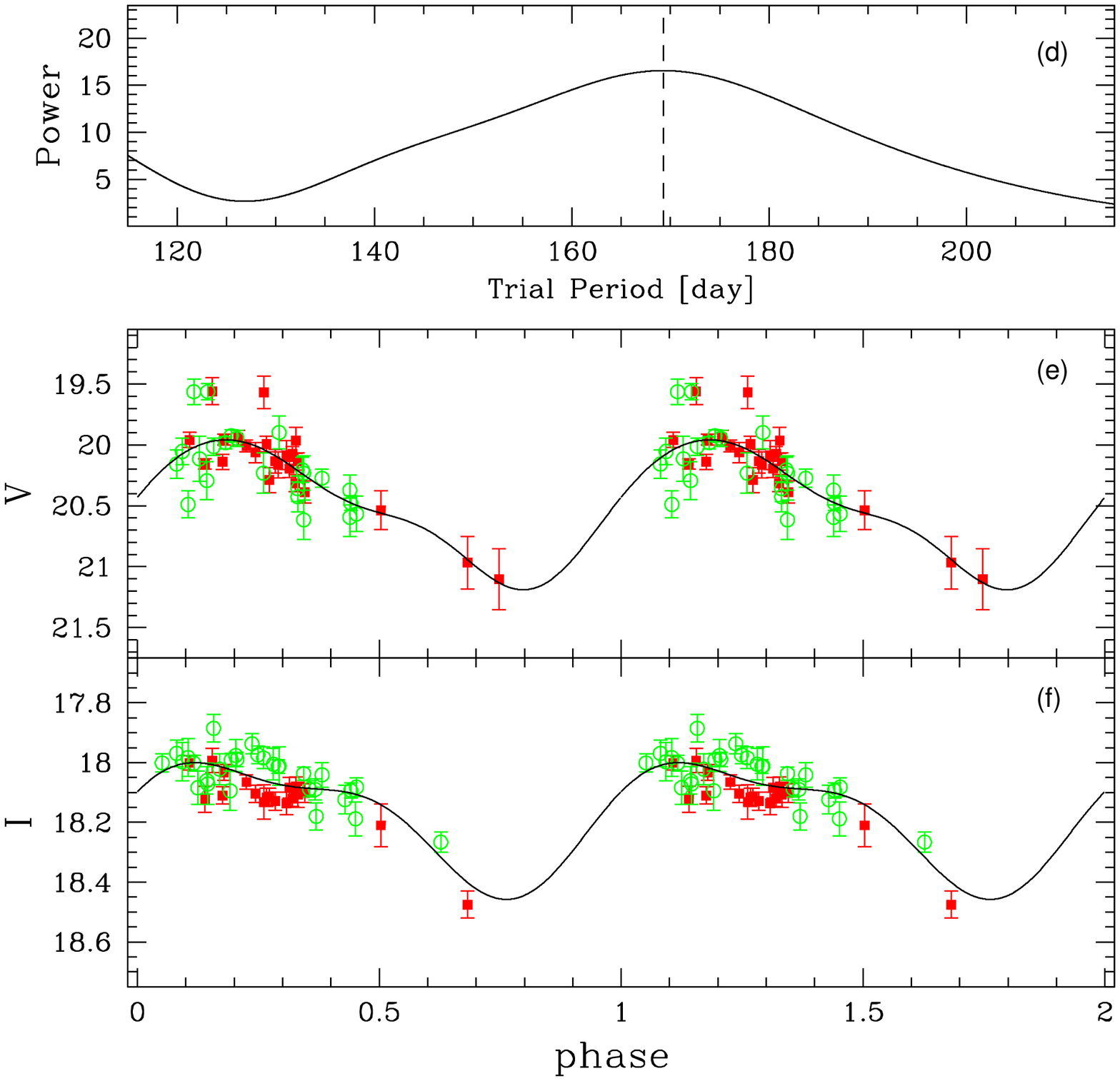} \\ 
    \includegraphics[angle=0,scale=0.40]{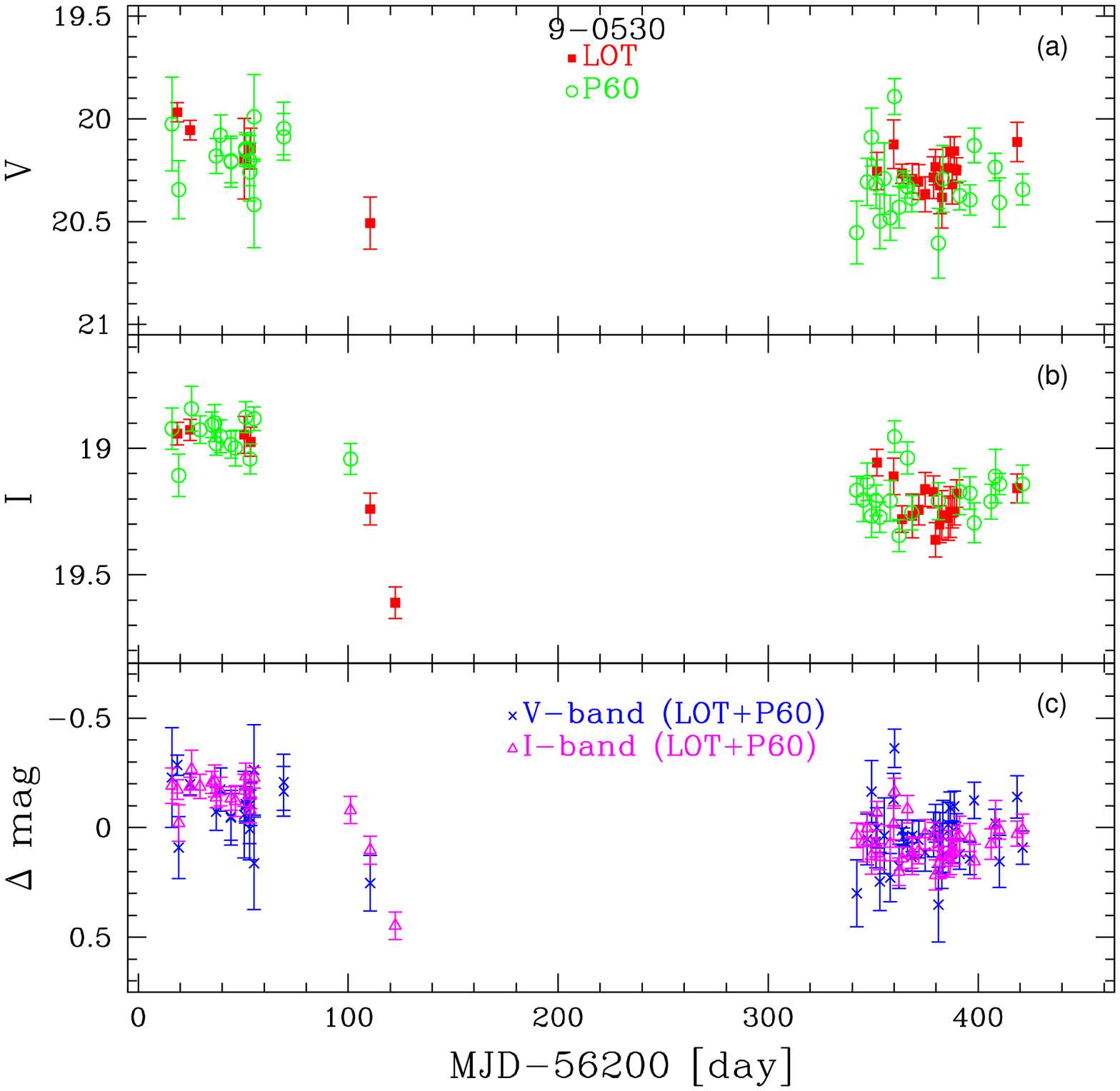} &
    \includegraphics[angle=0,scale=0.40]{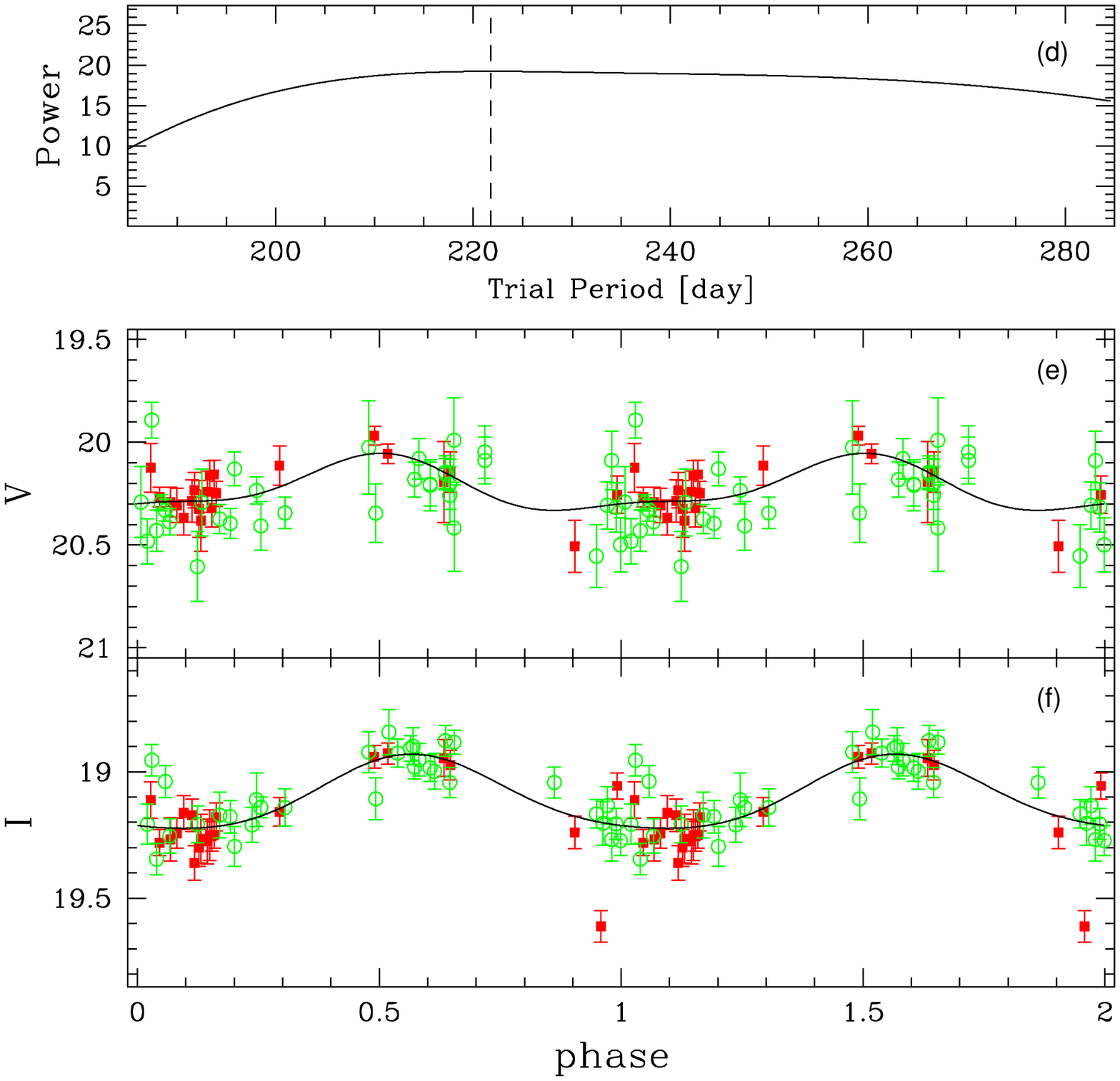} \\ 
  \end{array}$ 
  \caption{Same as Figure \ref{fig_80326}, but for candidate 4-1047, 8-0272 \& 9-0530.} \label{fig_badulpc1}
\end{figure*}

\begin{figure*}
  $\begin{array}{cc}
    \includegraphics[angle=0,scale=0.40]{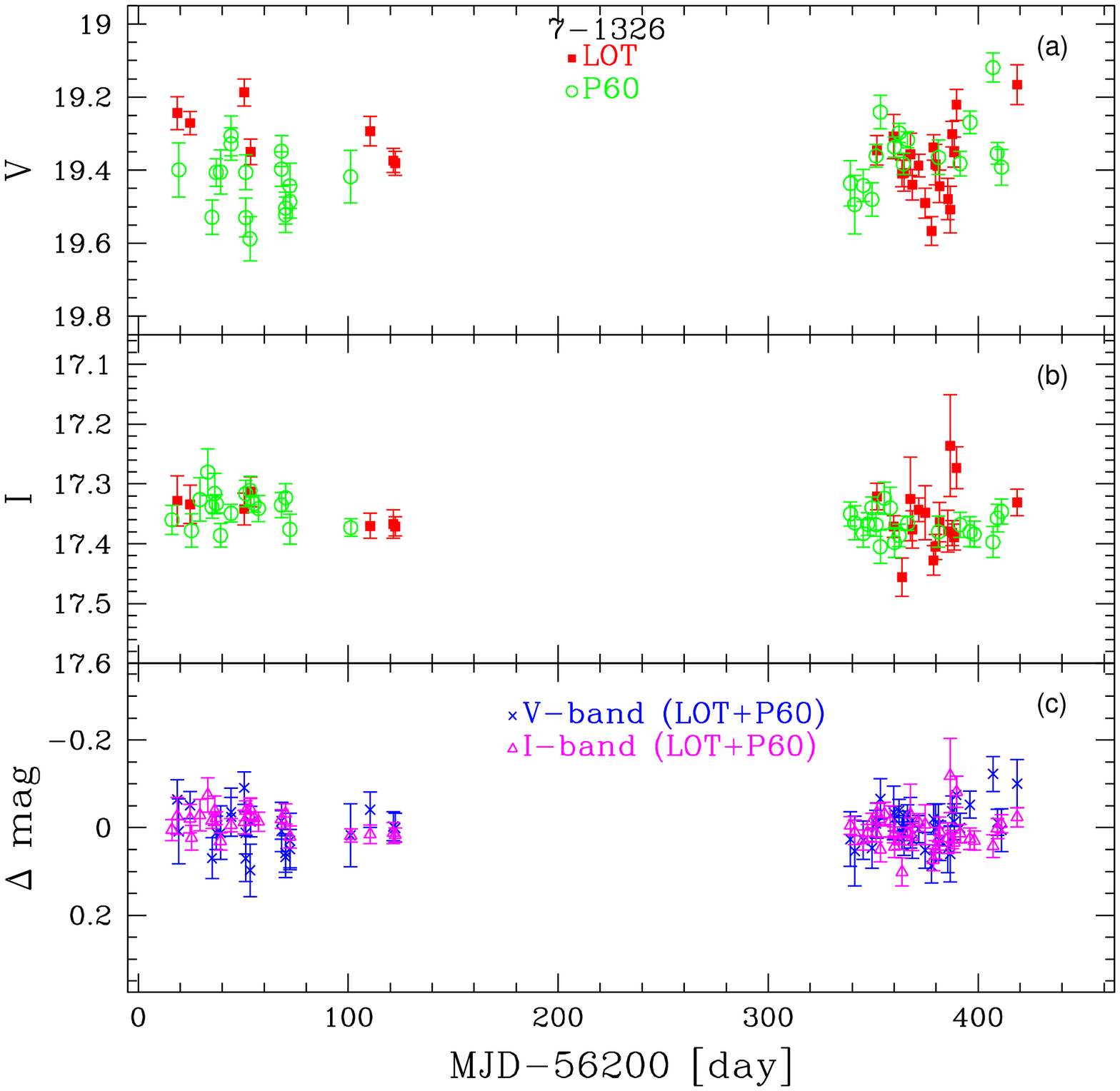} & 
    \includegraphics[angle=0,scale=0.40]{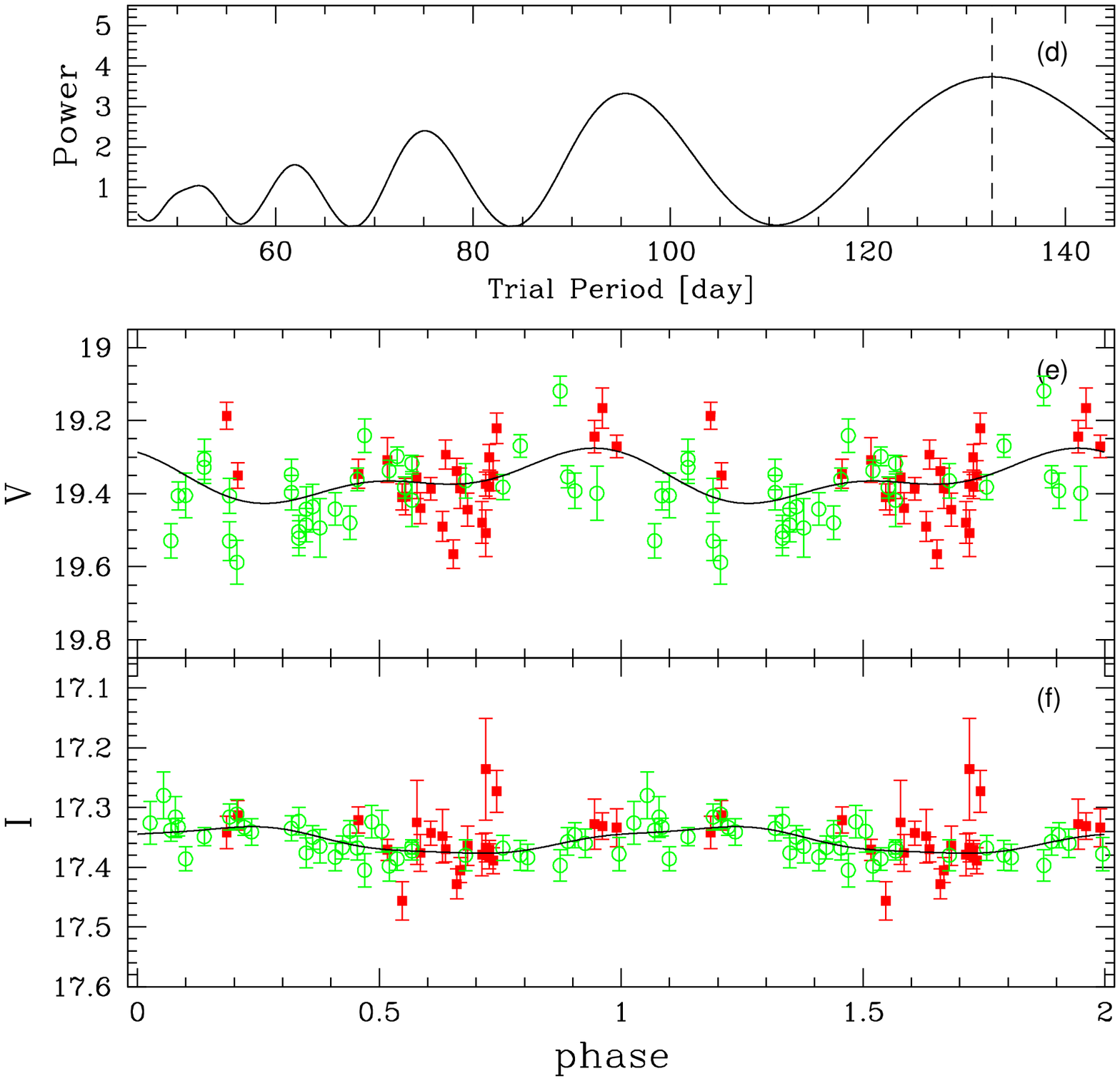} \\
    \includegraphics[angle=0,scale=0.40]{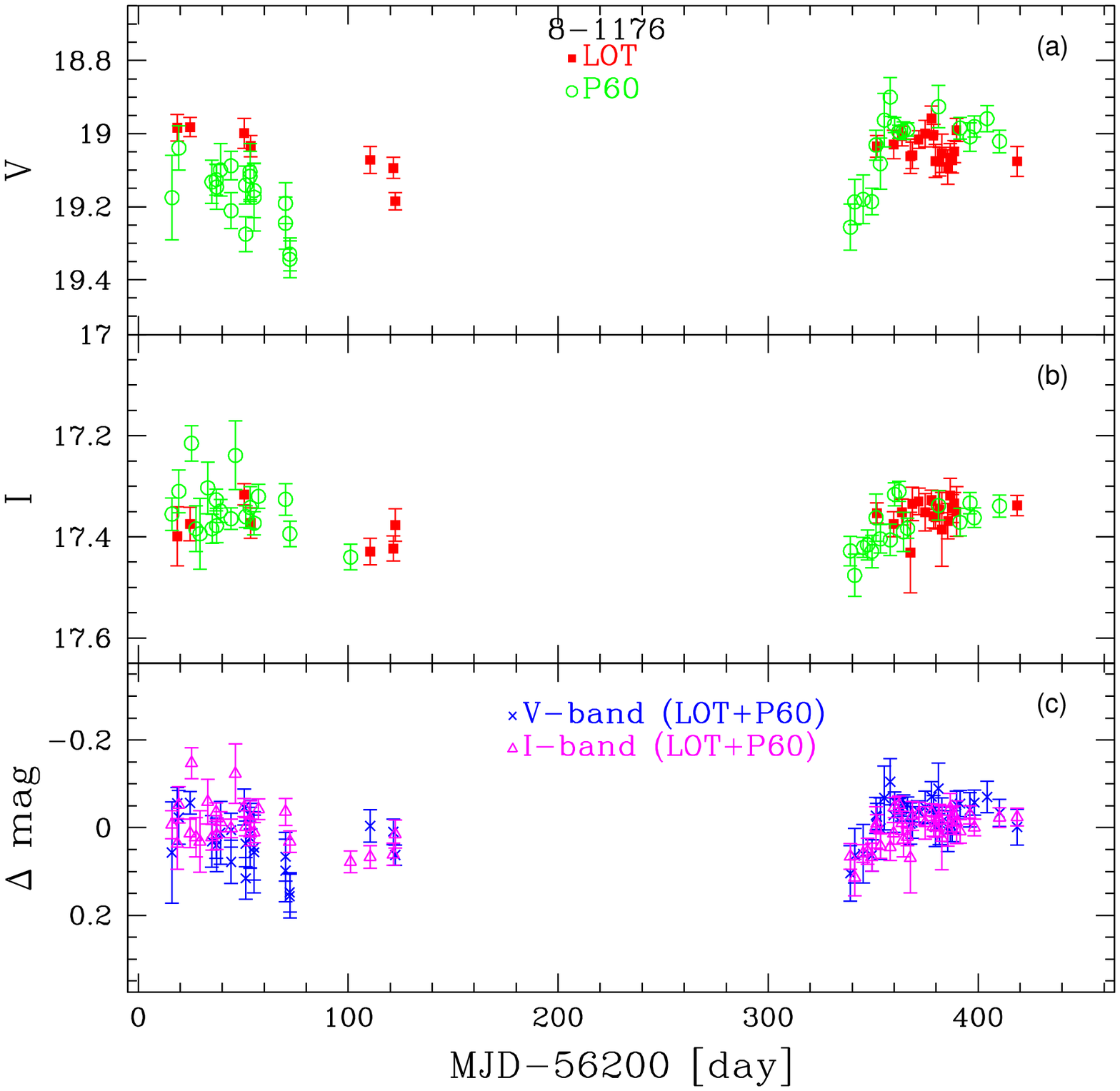} & 
    \includegraphics[angle=0,scale=0.40]{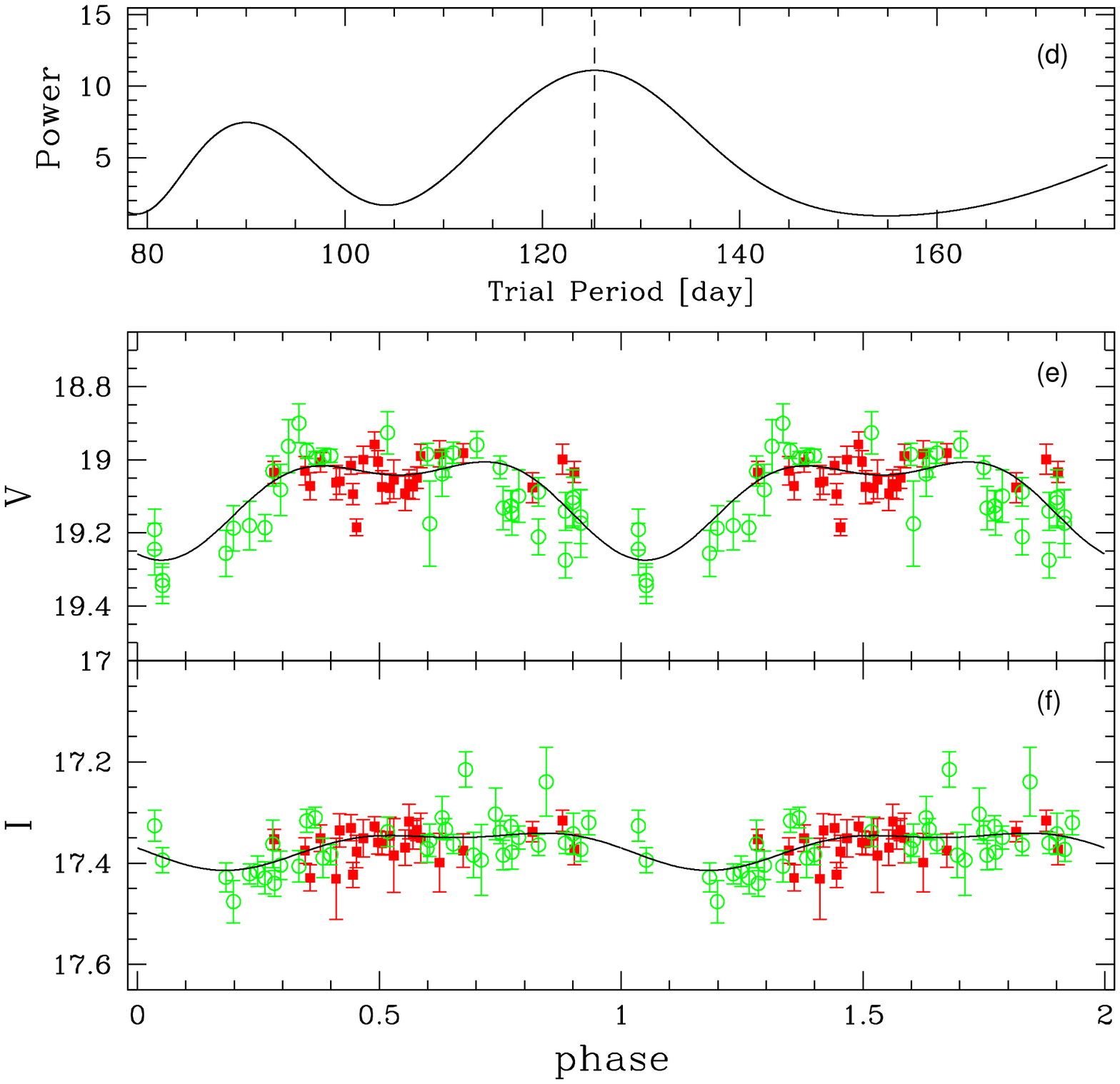} \\ 
    \includegraphics[angle=0,scale=0.40]{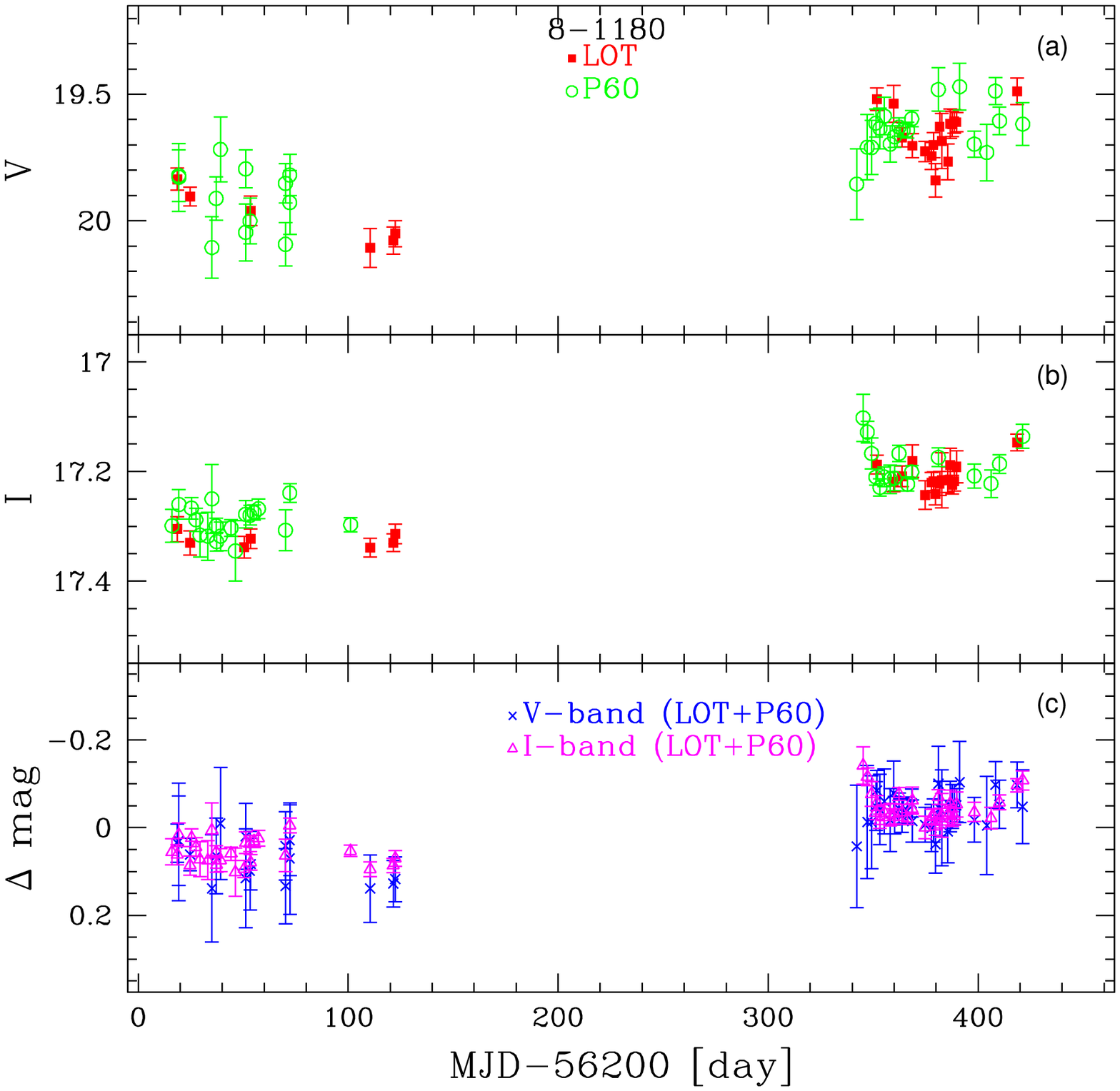} &
    \includegraphics[angle=0,scale=0.40]{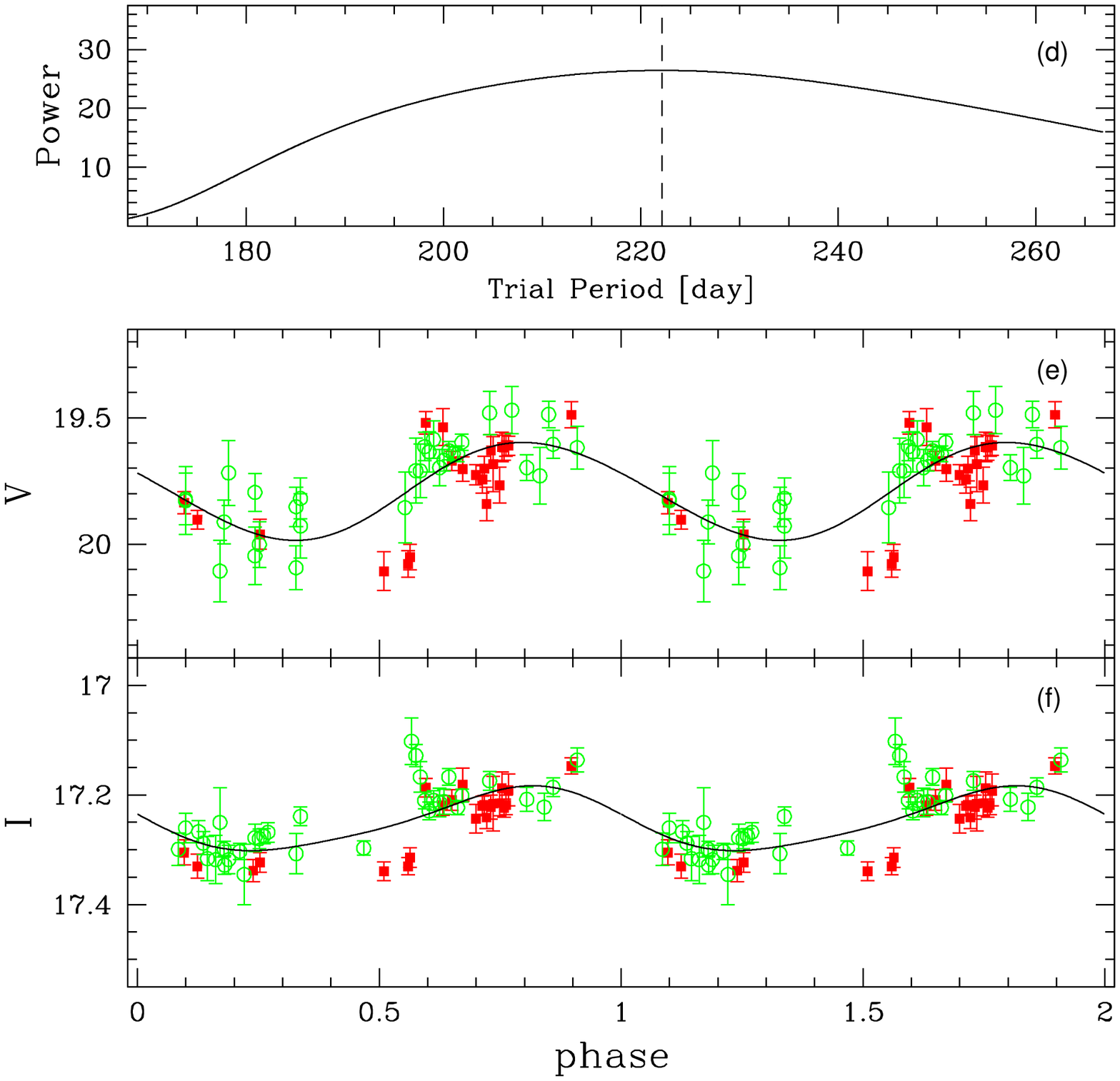} \\ 
  \end{array}$ 
  \caption{Same as Figure \ref{fig_80326}, but for candidate 7-1326, 8-1176 \& 8-1180. We found that an additional zero-point offset of $+0.078$, $+0.089$ and $+0.078$ is needed to be added to P60 data for these candidates with internal ID 7-1326, 8-1176 and 8-1180, respectively.} \label{fig_badulpc2}
\end{figure*}

\begin{deluxetable}{lllccc}
\tabletypesize{\scriptsize}
%\rotate
\tablecaption{$VI$-band Light Curves for the Candidates. \label{tab_lc}}
\tablewidth{0pt}
\tablehead{
\colhead{Candidate} & 
\colhead{Telescope} & 
\colhead{Band} &
\colhead{MJD} &
\colhead{$m$} &
\colhead{$\sigma_m$} 
}
\startdata
8-0326	& LOT	& V &	56218.62510 &	18.819 & 0.031 \\
8-0326	& LOT	& V &	56224.71680 &	18.931 & 0.034 \\
8-0326	& LOT	& V &	56224.71403 &	18.935 & 0.033 \\
8-0326	& LOT	& V &	56235.64163 &	19.026 & 0.038 \\
$\cdots$&$\cdots$&$\cdots$&$\cdots$&$\cdots$&$\cdots$
\enddata
\tablecomments{(This table is available in its entirety in machine-readable and Virtual Observatory (VO) forms.)}
\end{deluxetable}

\noindent where $w_i$ were the same weights as in equation (1), and $m_i^{\mathrm{cs,LGS}}$ were the $VI$-band magnitudes from the LGS catalog. The calibrated $VI$-band light curves for the eight candidates are given in Table \ref{tab_lc}, and displayed in panel (a) and (b) of Figures \ref{fig_80326}-\ref{fig_badulpc2}. 

\section{Analysis and Results}\label{sec_result}

\subsection{Period Search and Folded Light Curves}

The new $VI$-band light curves suggested that periods for the candidates need to be revised from Paper I. For each pair of the $VI$-band light curves, we first subtracted the arithmetic mean of the light curve, followed by normalization of the light curve using the peak-to-peak amplitude. We then applied the well-documented Lomb-Scargle algorithm to search for periodic signals by combining these mean-subtracted and normalized pair of $VI$-band light curves (as shown in panel (c) of Figures \ref{fig_80326}-\ref{fig_badulpc2}) to each of the candidates. We only searched for periodic signals around $\sim \pm30$ or $\sim \pm50$~days from the periods reported in Paper I (with the exception of candidate 8-0272, for which a shorter period range was searched for periodic signals). The periods corresponding to the highest peaks in the Lomb-Scargle periodogram were taken as the periods for these candidates. They are indicated by a vertical dashed line in panel (d) of Figures \ref{fig_80326}-\ref{fig_badulpc2}, as well as listed in the second column of Table \ref{tab_result}. We also ran a Monte-Carlo simulation to quantify the errors on the corresponding periodic signals. Assuming the photometric errors are Gaussian, about 10,000 simulated light curves were generated using the Box-Muller method for each of the candidates. The Lomb-Scargle algorithm was then applied to these light curves in a similar fashion to the combined $VI$-band light curves. The errors on the derived periods were then estimated as the rms of the corresponding peaks in the power spectra.

Using the periods given in Table \ref{tab_result}, we folded the $VI$-band light curves for these eight candidates. The folded $VI$-band light curves were then fitted with a low-order Fourier decomposition \cite[for example, see][]{simon1981} in the following form:

\begin{eqnarray}
m(\phi) & = & m_0 + \sum_{i=1}^{i=2,3,4} A_i \cos (2i \pi \phi + \Phi_i) 
\end{eqnarray}

\noindent where $\phi \in [0,1]$ were phases of the pulsating cycles. The fitted light curves, shown as black curves in panels (e) and (f) of Figures \ref{fig_80326}-\ref{fig_badulpc2}, were used to derive the $VI$-band mean magnitudes (by converting the fitted light curves to intensity, taking the average, and converting back to magnitude), as well as the extinction-free Wesenheit function $W$. In order to be consistent with previous works \citep{bird2009,fiorentino2012}, we adopted the same Wesenheit function in the form of $W=I-1.55(V-I)$ from \citet{udalski1999}. The coefficient of $1.55$ in the Wesenheit function is based on the extinction law from \citet{schlegel1998}. The derived mean magnitudes and the values of Wesenheit function are listed in Table \ref{tab_result}. 

The shape of the ULPC $VI$-band light curves are similar to the classical Cepheids \citep{bird2009,ngeow2013}, for which light curves exhibit the characteristic ``saw-tooth'' shape with a quick rise to maximum light followed by a slow decline to minimum light. Based on our $VI$-band light curves, only two of the candidates (8-0326 and 8-1498) exhibit Cepheid-like light curves, as shown in Figures \ref{fig_80326} and \ref{fig_81498}, respectively. The $VI$-band light curves for the other six candidates are given in Figures \ref{fig_badulpc1} and \ref{fig_badulpc2}. These curves do not display light curves that are expected for Cepheids. In the following sub-sections, the two candidates with Cepheid-like light curves are referred to as ``M31 ULPC candidates,'' while the other six candidates will be referred to as ``M31 non-ULPC candidates.'' Further confirmation or falsification of their ULPC nature will be detailed in the next sub-section. We note that the two ``M31 ULPC candidates'' have well-determined periods (with errors of the order of $\sim0.2\%$), while the periods for the ``M31 non-ULPC candidates,'' with the exception of the candidate with an internal ID of 4-1047, have a much larger error (from $\sim2.6\%$ to $\sim8\%$). Furthermore, as mentioned in Paper I, these two ``M31 ULPC candidates'' have previously been classified as Cepheids in the literature \citep[for example, in][but they have incorrect periods]{magnier1997}.

\begin{deluxetable}{lcccc}
\tabletypesize{\scriptsize}
%\rotate
\tablecaption{Revised Periods and Mean Magnitudes for the Candidates. \label{tab_result}}
\tablewidth{0pt}
\tablehead{
\colhead{Candidate} & 
\colhead{Period (days)} & 
\colhead{$<V>$} &
\colhead{$<I>$} &
\colhead{$W$} 
}
\startdata
\cutinhead{M31 ULPC Candidates}
8-0326  & $74.427\pm0.120$  & 18.684  & 17.256 & 15.043 \\
8-1498  & $83.181\pm0.178$  & 18.856  & 17.783 & 16.120 \\
\cutinhead{M31 non-ULPC Candidates}
8-1176  &$125.313\pm4.832$  & 19.095  & 17.367 & 14.689 \\
7-1326  &$132.608\pm4.543$  & 19.357  & 17.356 & 14.254 \\
8-0272  &$169.262\pm4.461$  & 20.464  & 18.176 & 14.630 \\
4-1047  &$216.169\pm0.907$  & 18.625  & 17.548 & 15.879 \\
9-0530  &$221.779\pm17.695$ & 20.218  & 19.092 & 17.347 \\
8-1180  &$222.124\pm5.959$  & 19.780  & 17.246 & 13.318 
\enddata
\end{deluxetable}

The two M31 ULPC candidates have periods of $\sim74.4$ and $\sim83.2$~days, respectively. According to the definition by \citet{bird2009}, a ULPC should have a period longer than $80$~days. Hence, the $\sim74.4$~days ULPC found in this work may not be classified as ULPC. However, an $80$~day period cut is rather arbitrary, and we prefer to identify the ULPCs based on the discussion given in Section 3.2. Furthermore, initial period detection searches for this candidate gave results in excess of $80$~days in Paper I, and was refined to a shorter period with other period-search algorithms and $VI$-band light curve data. Therefore, we retained this candidate in the sample of M31 ULPCs. 

\begin{figure*}[!t]
  \plottwo{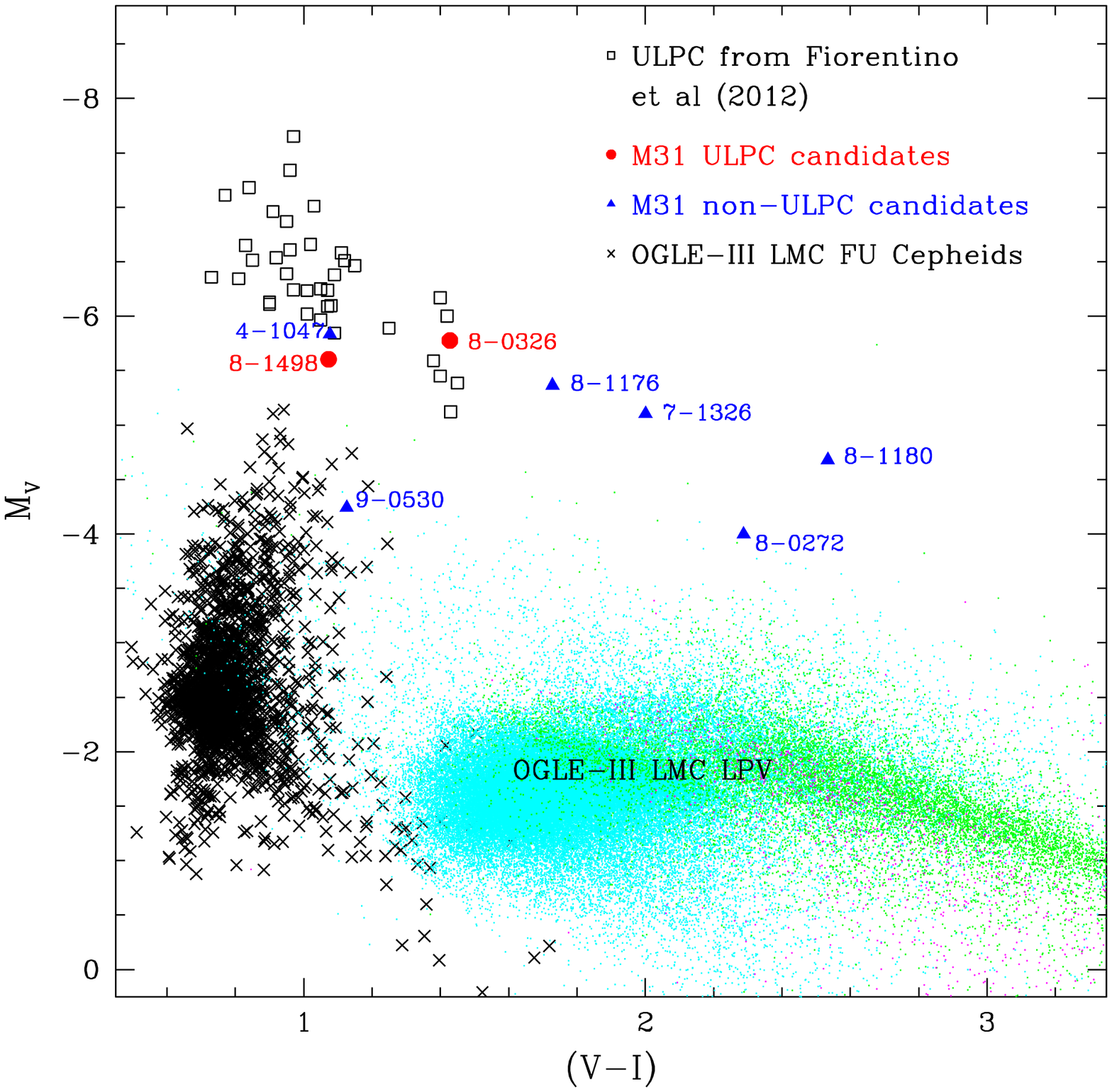}{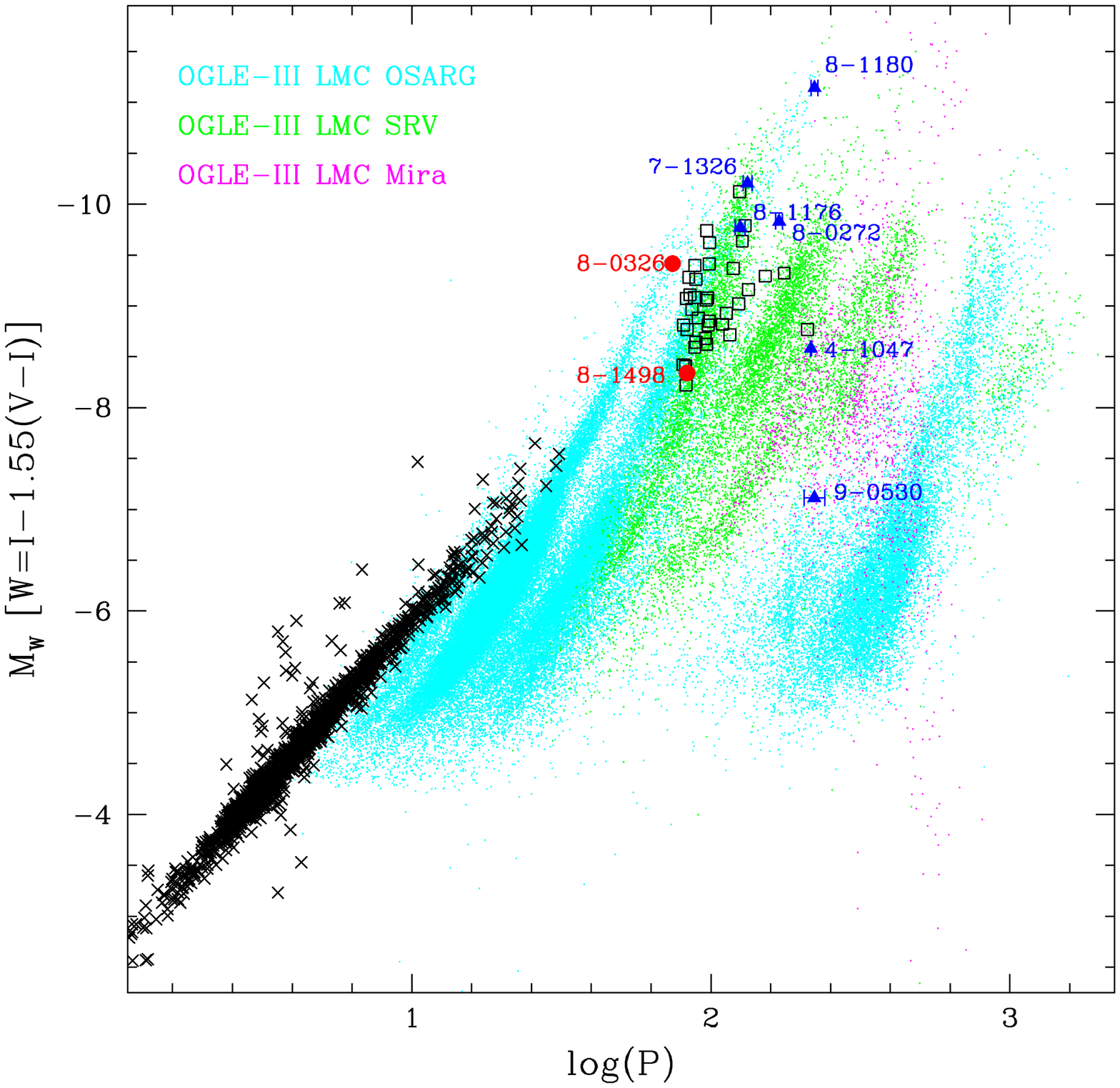}
  \caption{Comparison of our eight candidates to the known ULPCs listed in \citet{fiorentino2012}, as well as the LMC Cepheids and long-period variables (LPV) from the OGLE-III database, in the observed CMD (left panel) and PWD (right panel). We included LPVs to aid in falsifying the ULPC nature of our candidates. The LPVs were classified as small amplitude red giants (OSARG), semi-regular variables (SRV), and Mira by the OGLE team \citep{soszynski2009}. Note that the saturation limit in OGLE-III data is around $\sim13$~mag to $\sim13.5$~mag. This translates to $M_V\sim -5.0$~mag to $\sim -5.5$~mag.}
  \label{fig.cmdpl}
\end{figure*}

\begin{figure*}[!t]
  \plottwo{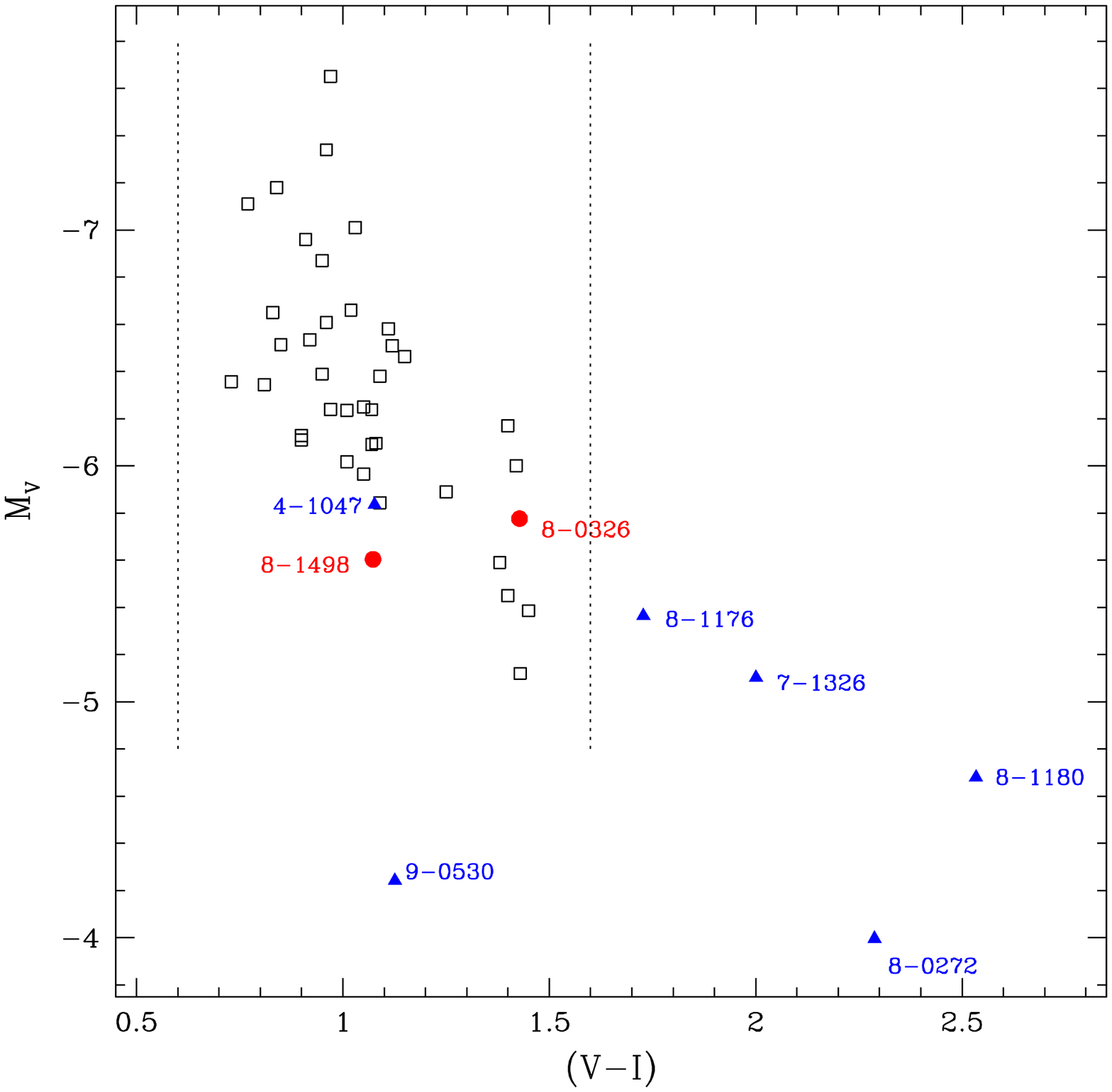}{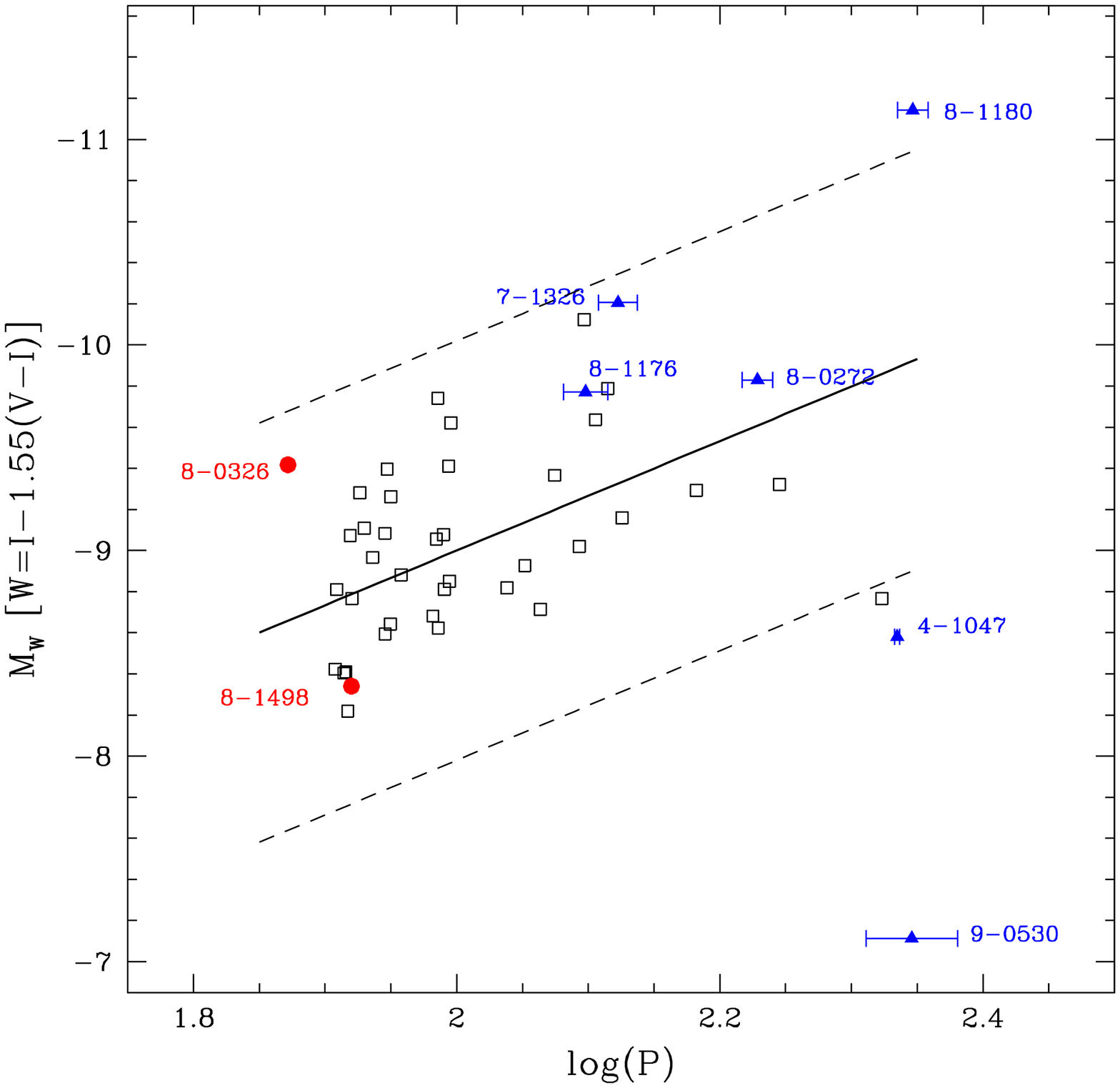}
  \caption{Observed CMD (left panel) and PWD (right panel) only showing the ULPCs. Since the predicted boundary of the ULPC instability strip from theoretical calculations is not yet available, we used the color range available from known ULPCs \citep{fiorentino2012}, $0.6<(V-I)<1.6$, to represent the (crude) boundary of the instability strip (shown as vertical lines in the left panel). The solid line in the right panel is the Period-Wesenheit relation adopted from \citet{fiorentino2012}, and the dashed lines represent the $\pm3\sigma$ boundary (where $\sigma=0.34$~mag).}
  \label{fig.enlarge}
\end{figure*}

\subsection{Verification of the ULPC Candidates}

Besides the shapes of the $VI$-band light curves, we verify/falsify the ULPC nature of our eight candidates by comparing them to the known ULPCs \citep[given in][]{fiorentino2012} in the color-magnitude diagram (CMD) and the Period-Wesenheit diagram (PWD), collectively presented in Figure \ref{fig.cmdpl}. Using the CMD and PWD to identify Cepheids has been done, for example, in \citet{udalski1999b} and \citet{kodric2013}. Only those Cepheids identified with periods longer than $\sim80$~days will be re-classified as ULPC \citep{bird2009,fiorentino2012}. When making such comparisons, we have adopted the distance modulus of M31 as $\mu_{M31}=24.46$~mag, recommended by \citet{degrijs2014b}, to convert the observed magnitudes of these candidates to the absolute magnitudes. We also included the LMC FU Cepheids and long-period variables (LPV) based on the OGLE-III database, taken from \citet{soszynski2008} and \citet{soszynski2009} respectively, in these diagrams. Again, we adopted $\mu_{LMC}=18.49$~mag \citep{degrijs2014a} when converting the observed magnitudes to absolute magnitudes. We enlarged Figure \ref{fig.cmdpl} by only showing the known ULPCs and our eight candidates in Figure \ref{fig.enlarge}. Figures \ref{fig.cmdpl} and \ref{fig.enlarge} reveal that the two ``M31 ULPC candidates'' are indeed ULPC, as both of them fall in the parameter spaces defined by the known ULPCs. The remaining six ``M31 non-ULPC candidates'' are falsified as ULPC based on their location in the CMD and PWD diagrams, even though a few of them may be consistent with ULPCs in either the CMD or PWD diagram, but not in both. For example, candidates 4-1047 and 9-0530 fall within the ULPC instability strip, but they are outside the $\pm3\sigma$ boundary of the Period-Wesenheit (PW) relation. Similarly, candidates 7-1326, 8-0272 and 8-1176 are located within the $\pm3\sigma$ boundary of the PW relation, however they are too red in the CMD to be considered as ULPC. Since the main goal of this work is to verify the ULPC nature of our candidates in M31, determining the nature of these six non-ULPC candidates is beyond the scope of this paper, and they will not be discussed further.

\begin{deluxetable*}{lcccl}
\tabletypesize{\scriptsize}
%\rotate
\tablecaption{M31 Cepheids with Periods Greater than $\sim75$~days but Less than $80$~days. \label{tab_more ulpc}}
\tablewidth{0pt}
\tablehead{
\colhead{Name or ID} & 
\colhead{Period (days)} & 
\colhead{$\alpha (J2000)$} &
\colhead{$\delta (J2000)$} &
\colhead{Reference} 
}
\startdata
vn.4.2.678          & 78.00 & 11.06420 & 41.56927 & \citet{riess2012} \\
PSO J010.5806$+$40.8319 & 74.79 & 10.58066 & 40.83192 & \citet[][only for Cepheids classified as ``FU'']{kodric2013}  
\enddata
\end{deluxetable*}

It is expected that a galaxy would have small number of ULPCs for two reasons. Firstly, ULPCs are intermediate- to high-mass stars with masses between $\sim15M_\odot$ ang $\sim20M_\odot$ \citep{bird2009,fiorentino2012}. Hence, their number should be lower than their shorter period counterparts that have lower mass. Second, these intermediate- to high-mass stars need to cross the instability strip in order to pulsate and the time that they spend inside the instability strip is relatively short as compared to their evolutionary age. Based on the list compiled in \citet{fiorentino2012}, the number of ULPCs in a given galaxy ranges from one (NGC 6822), two (M81 and I Zw 18), three (SMC ang NGC 300), four (LMC, NGC 1309 and NGC 3021), five (NGC 55), and nine (NGC 3370). Then, it is not a surprise that only two ULPCs were found in M31. Another reason for detecting a small number of ULPCs in M31 could be due to characteristics of PTF data. For example, the one CCD that is out of commission covers a portion of M31's disk (see Figure \ref{fig.surveys}), and the stars in M31's bulge region are not resolvable in PTF images. Further discussion on the effect of PTF data can be found in Paper I, and will not be repeated here. A few additional ULPCs might be discovered from the three years of the Pan-STARRS1's PAndromeda Survey (M. Kodric et al. 2014, private communication). On the other hand, if the period cut of $\sim80$~days that defines ULPC is lowered to $\sim75$~days, then there are a few more very long-period Cepheids in M31 that could be classified as ULPCs. These are listed in Table \ref{tab_more ulpc}. They were not reported in Paper I because their periods are lower than the search criterion of $P>80$~days. Furthermore, there is no $VI$-band photometric data available for these Cepheids currently.

\subsection{Distance Scale Application}

In the previous sub-section, we adopted a distance modulus for M31 in order to verify the ULPC nature of our candidates. In this sub-section, we reverse the problem by using the two confirmed ULPCs to determine the distance modulus of M31, i.e., $\mu = W - M_W$. The absolute magnitude $M_W$ of the two ULPCs can be determined from the latest PW relation as given in \citet{fiorentino2012}: $M_W = -2.66 \log(P) - 3.68$, with a dispersion of $\sigma = 0.34$. Since the PW relation is a statistical relation, the dispersion of this relation ($\sigma$) will dominate the error term of the derived distance modulus for individual ULPCs. Hence, we have adopted the dispersion of the PW relation as the error in the calculated distance moduli for the two M31 ULPCs: $\mu$(8-0326) $= 23.70\pm0.34$~mag and $\mu$(8-1498) $= 24.91\pm0.34$~mag. Since the weights for these two distance moduli are the same, taking an average of them reduces to the case of an unweighted mean. This procedure yielded $\mu_{M31,ULPC}=24.30\pm0.76$~mag, where the error on the averaged distance modulus is calculated using the small number statistics given in \citet[][p. 202]{keeping1962}. The large error on the determined distance modulus is mainly due to the combination of small number statistics (as there are only two ULPCs found in M31) and the large dispersion of the PW relation for ULPCs. 

G. Fiorentino (2014, private communication) suggested using the PW relation derived from ULPCs in metal-rich galaxies (hereafter metal-rich ULPCs). However, the sample of metal-rich ULPCs given in \citet{fiorentino2012}, i.e. those with $12+\log [O/H]>8.4$~dex, only occupied a narrow range in $\log (P)$. This causes the derived PW relation to be unreliable. Instead, we have averaged the absolute Wesenheit magnitudes for this sample of metal-rich ULPCs, yielding $\overline{M_W}=-8.84$~mag. Using this absolute Wesenheit magnitude, the derived distance modulus to M31 with our two ULPCs is $\mu_{M31,ULPC}=24.42\pm0.68$~mag with a (unbiased) standard deviation of $0.76$. The value of this distance modulus is almost identical to the value recommended in \citet{degrijs2014b}. Note that the sample of metal-poor ULPCs ($12+\log [O/H]<8.4$~dex) in \citet{fiorentino2012} is the same as the sample given in \citet{bird2009}. The derived distance modulus is $\mu_{M31,ULPC}=24.74\pm0.68$~mag when using the PW relation given in \citet{bird2009}.

\begin{figure}
  \plotone{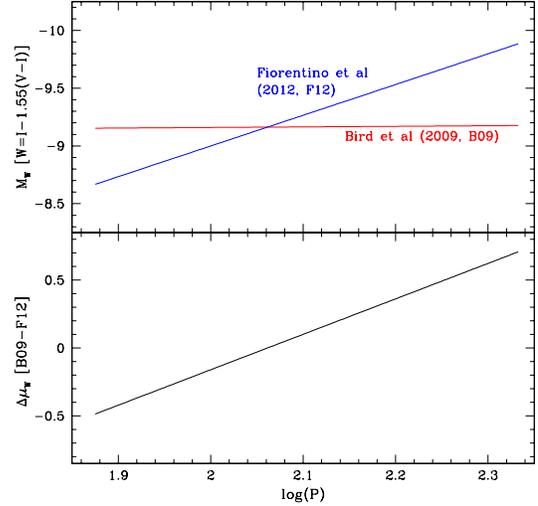}
  \caption{{\bf Top:} comparison of the two Period-Wesenheit relations adopted from \citet{bird2009} and \citet{fiorentino2012}; {\bf Bottom:} difference in the derived distance modulus from these two Period-Wesenheit relations as a function of the ULPC periods.}
  \label{fig_comparepw}
\end{figure}

\begin{figure}
  \plotone{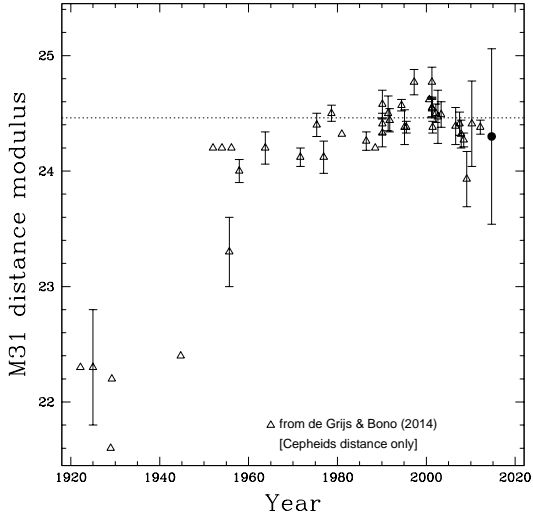}
  \caption{Comparison of the M31 distance modulus based on the two ULPCs in this work ($\mu_{M31,ULPC}=24.30\pm0.76$~mag, filled circle) to other distance moduli determined using Cepheids \citep[as compiled in][open triangles]{degrijs2014b}. Note that some of the distance moduli, especially those in early years, did not include an error estimation. The horizontal dashed line represents the recommended distance modulus of $24.46$~mag \citep{degrijs2014b}.}
  \label{fig.m31dm}
\end{figure}

The distance moduli derived using the PW relations given in \citet{fiorentino2012} and \citet{bird2009} differed by $0.44$~mag, albeit with large uncertainties. This reflects one of the current problems in using ULPCs to derive distances: the calibration of the ULPC PW relations is still uncertain with many discrepant results. The slope of the PW relation derived in \citet{bird2009}, based on 18 ULPCs in metal-poor host galaxies, is essentially zero but with a large error ($-0.05\pm0.54$). In contrast, the PW slope given in \citet{fiorentino2012} is $-2.66$ (though no error on the slope is quoted in their paper). The intercepts of these two Period-Wesenheit relations even displayed a larger difference: $-9.06\pm1.12$ versus $-3.68$. The top panel in Figure \ref{fig_comparepw} compares these two PW relations, while the bottom panel of the same figure shows the difference of the derived distance modulus based on these two PW relations: this can be as large as $\sim0.5$~mag. This disagreement between the two PW relations suggests more work is needed in the future to properly calibrate the ULPC PW relation. Nevertheless, the distance modulus derived from the \citet{fiorentino2012} PW relation will be adopted in this work, because this Period-Wesenheit relation is derived from a larger sample of ULPCs than can be found in the literature. Figure \ref{fig.m31dm} shows a comparison of the adopted distance modulus (based on the two M31 ULPCs) to other distance moduli given in literature. Our calculated distance modulus is consistent with some of recent determinations, including the recommended value of $24.46\pm0.10$~mag \citep{degrijs2014b}.

\begin{figure}
  \plotone{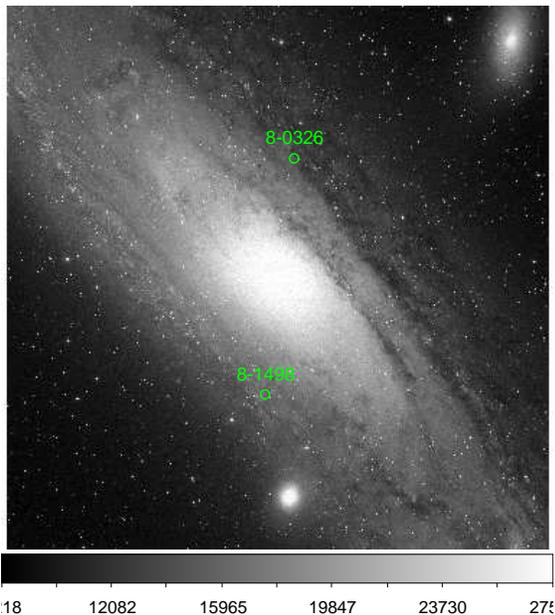}
  \caption{Locations of the two ULPCs in M31. The M31 image, with a size of $60'\times 60'$, was downloaded from the Digitized Sky Surveys (DSS) archive.}
  \label{fig.closeup}
\end{figure}

\begin{figure*}[!t]
  \plottwo{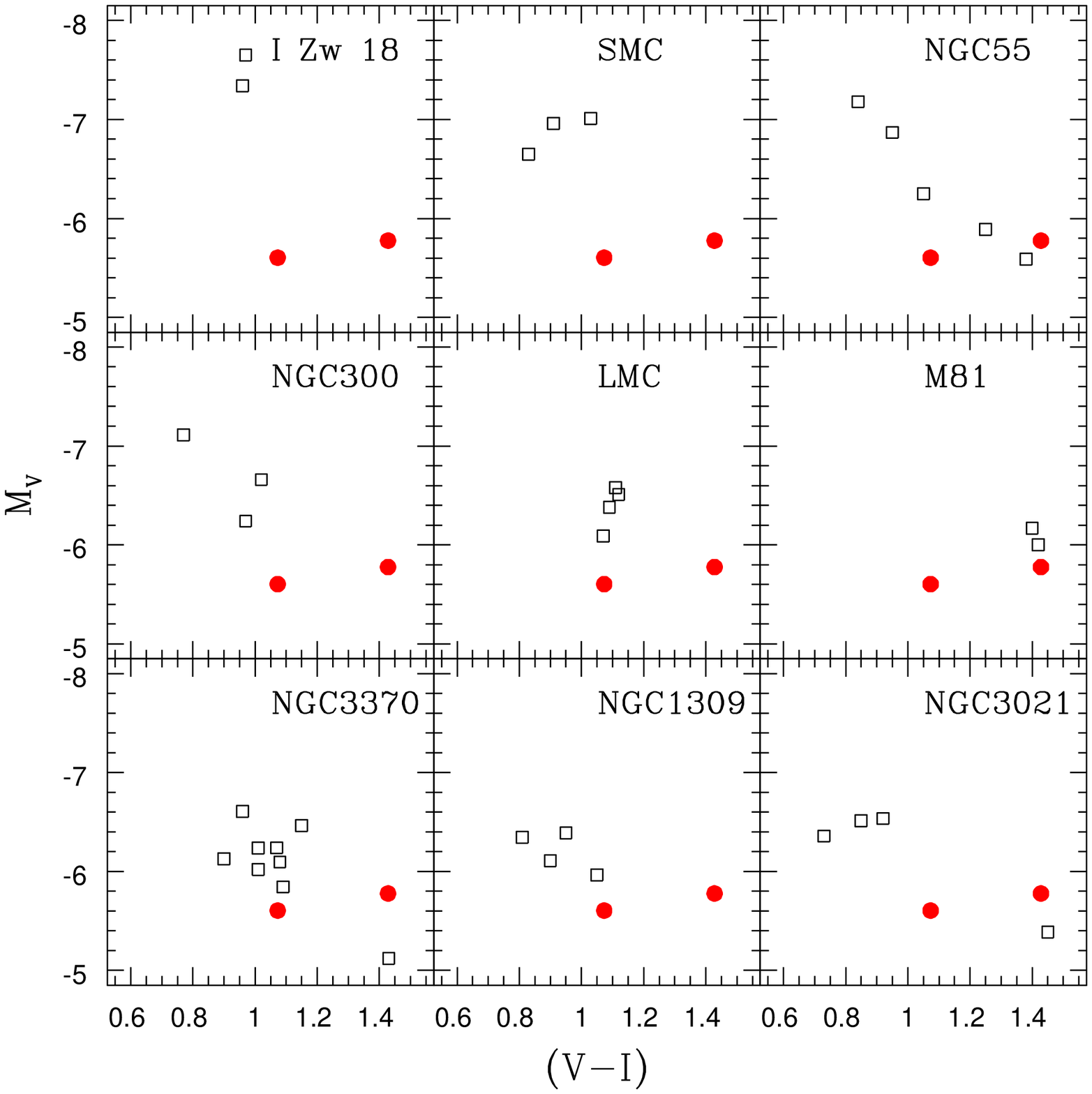}{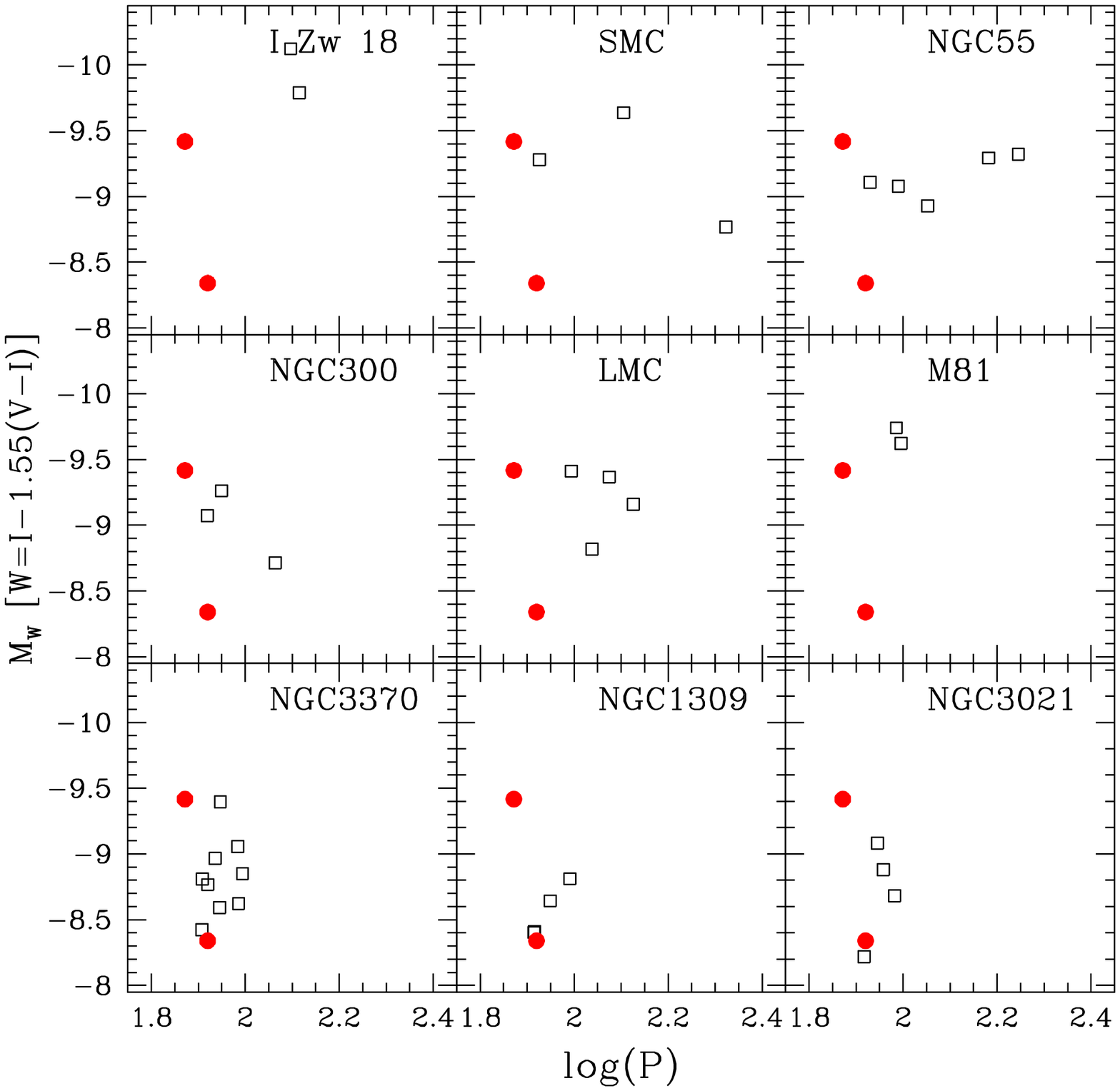}
  \caption{Comparison of the locations of the two M31 ULPCs (red filled circles) with ULPCs in other galaxies \citep[open squares, taken from][]{fiorentino2012} in the CMD (left panel) and PWD (right panel). We excluded the ULPC in NGC 6822, as this galaxy has only one ULPC. Note that sub-figures in each panel are ordered according to the metallicity of the host galaxy, which is given in \citet{fiorentino2012}.}
  \label{fig_indgal}
\end{figure*}

We note that the Wesenheit magnitudes and the distance moduli for these two ULPCs differ by $\sim1$~mag. This is not due to the ``depth effect,'' as one ULPC is located at the near-side and another ULPC is located at the far-side of M31 (see Figure \ref{fig.closeup}). Assuming the size of M31 is about 40~kpc, the maximum difference in distance modulus for two stars located at the two extreme edges is $\sim0.11$~mag. This is $\sim10\times$ smaller than the observed difference. Instead, the observed $\sim1$~mag difference in Wesenheit magnitudes or distance moduli for these two ULPCs is consistent with the spread of $M_W$ at $\log(P)\sim 1.9$, as shown in right panel of Figure \ref{fig.cmdpl} \citep[or Figure 2 in][]{fiorentino2012}, and the expected spread of the PW relation ($4\times \sigma = 1.36$~mag). We believe this is due to their relative locations within the instability strip on the CMD, depending on their evolutionary status.\footnote{The evolutionary status of ULPCs could be affected by various physical parameters, such as age, metallicity, mass-loss, and etc. However, a detailed investigation of their evolutionary status is beyond the scope of this paper.} Figure \ref{fig_indgal} compares the locations of the ULPCs in individual galaxies, as listed in \citet{fiorentino2012}, to the locations of the two M31 ULPCs in CMD and PWD. This shows that ULPCs in some galaxies (such as NGC 3370 and NGC 3021) exhibit a similar spread in Wesenheit magnitudes as in the case of the M31 ULPCs. The spread of Wesenheit magnitudes, at a given period, can be expressed as $\Delta M_W = \Delta M_I - 1.55\Delta (V-I)$, where $\Delta M_I$ and $\Delta (V-I)$ represent the range of magnitudes and colors, respectively, for ULPCs in the same galaxy. Therefore, assuming $\Delta M_I\sim 0$ and a difference of $\Delta (V-I)\sim 0.5$~mag could translate to a difference of $\Delta M_W \sim 0.8$~mag, which is close to the observed spread of M31 ULPCs.\footnote{To be more specific, using the values give in Table \ref{tab_result}, we found that $\Delta I=-0.527$~mag and $\Delta (V-I)=0.355$~mag for the two M31 ULPCs yielding $\Delta M_W\sim1.08$~mag as observed.} This implies that at a given period, the Wesenheit magnitudes or distance moduli for the ULPCs can differ by as much as by $\sim1$ magnitude. This calls into question the use of ULPCs as standard candles.

\section{Conclusion}\label{sec_conclusion}

In this work, we have presented the $VI$-band follow-up observations of eight ULPC candidates in M31. Based on their light curve shapes and their locations on the CMD and PWD, we verified that two of the candidates are indeed ULPCs, and the remaining six candidates are not ULPCs (and possibly belong to the LPV class). These six non-ULPC candidates also showed that other types of LPV could be mis-classified as ULPCs (and vice versa) if they have not had appropriate follow-up observations in the $VI$-bands \citep[see also][]{ngeow2013}. We then used these two confirmed ULPCs to test their applicability in distance scale work by deriving the distance modulus to M31. Our derived distance modulus is consistent with the recommended value given in the literature, but with a large error of $0.76$~mag. We have demonstrated three problems when using ULPCs as a distance indicator.

\begin{enumerate}
\item Small number statistics: the number of ULPCs is expected to be small in a host galaxy; hence, the ULPCs do not sample the instability strip well.

\item The discrepancy between PW relations in the literature: the two available PW relations from \citet{bird2009} and \citet{fiorentino2012} are in disagreement. As a result, the derived distance moduli can differ by as much as $\sim0.5$~mag. 

\item The large dispersion of the PW relation: the current calibration of PW relation, defined as $W=I-1.55(V-I)$, for ULPCs still possesses a large dispersion compared to their shorter period counterparts \citep[which is $\sim 5 \times$ smaller as shown in][]{ngeow2009}. Furthermore, at a given period, the back-to-back scatter in the PWD can be as large as $\sim 1$~mag.
\end{enumerate}

\noindent Combining these three problems with currently available PW relations results in a less precise and accurate distance modulus. As mentioned in Fiorentino's oral presentation\footnote{At the 2014 ``Extra-Galactic Distance Scale'' Workshop.}, ULPCs are not yet ready to be used as a standard candle. Larger samples of ULPCs with well calibrated distance host galaxy distances are needed in order to derive a better PW relation (with smaller $\sigma$) before using  ULPCs in ``one-step'' determinations of the Hubble constant in the future.

\begin{figure*}
  $\begin{array}{ccc}
    \includegraphics[angle=0,scale=0.28]{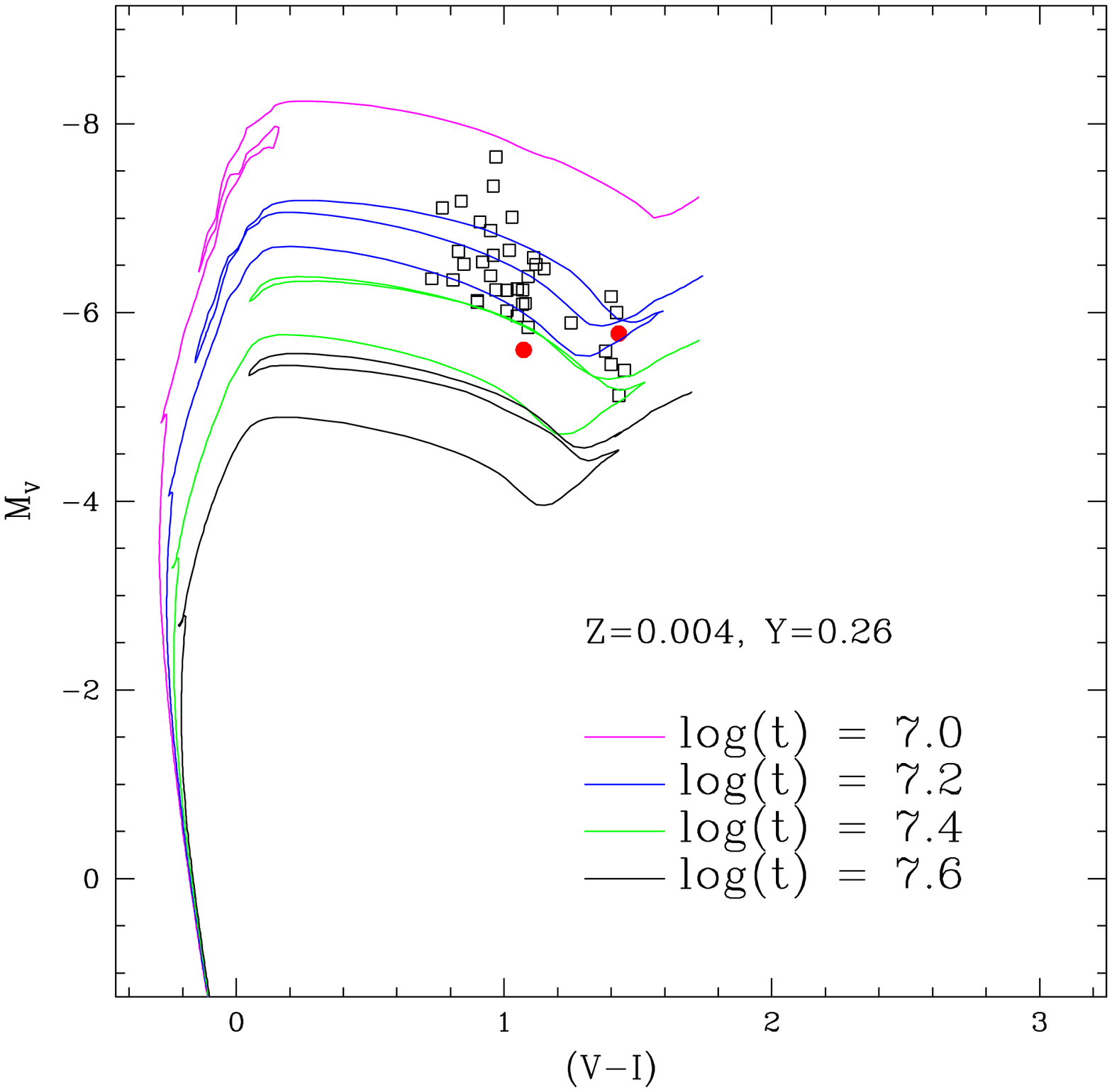} & 
    \includegraphics[angle=0,scale=0.28]{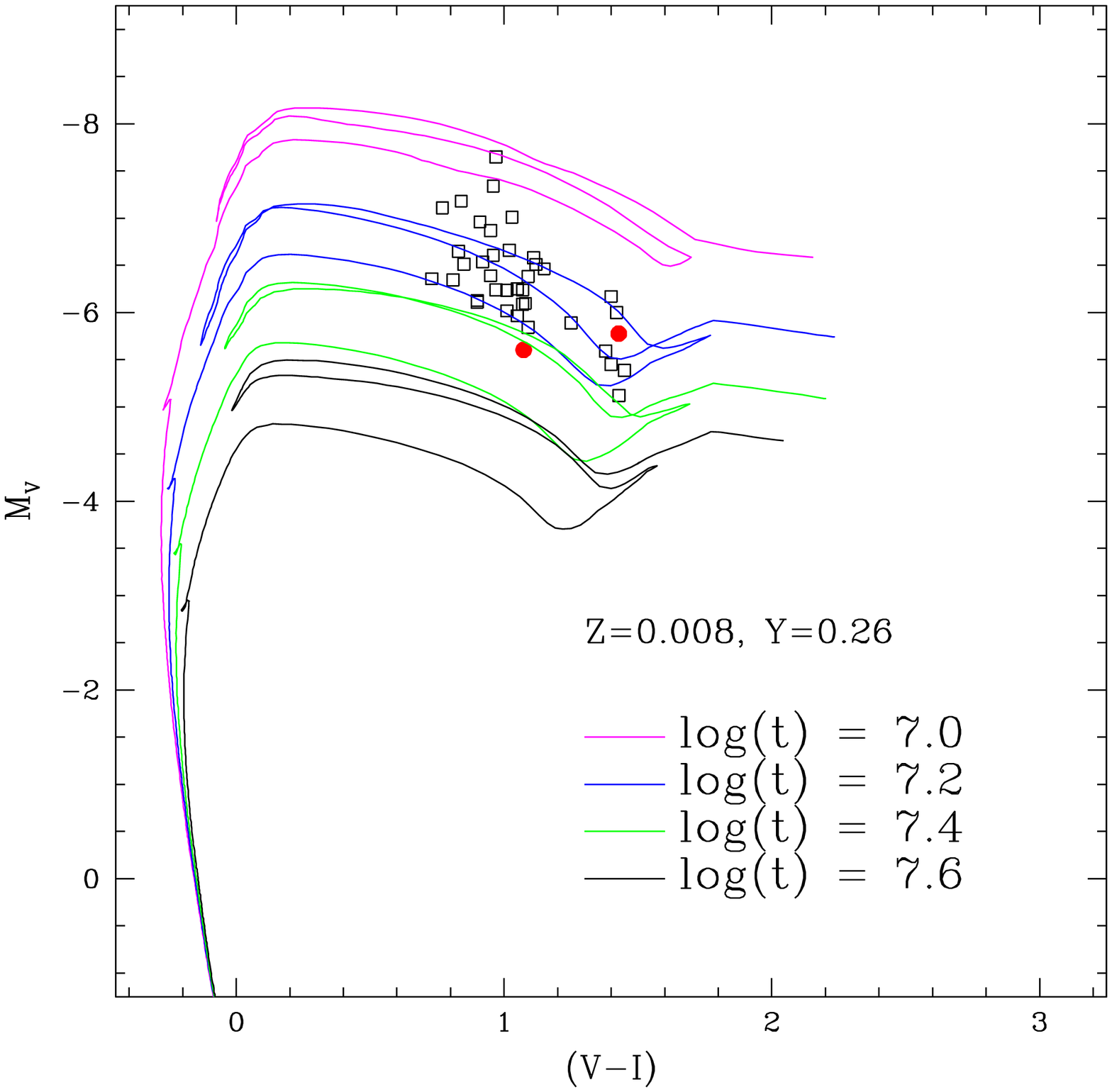} &
    \includegraphics[angle=0,scale=0.28]{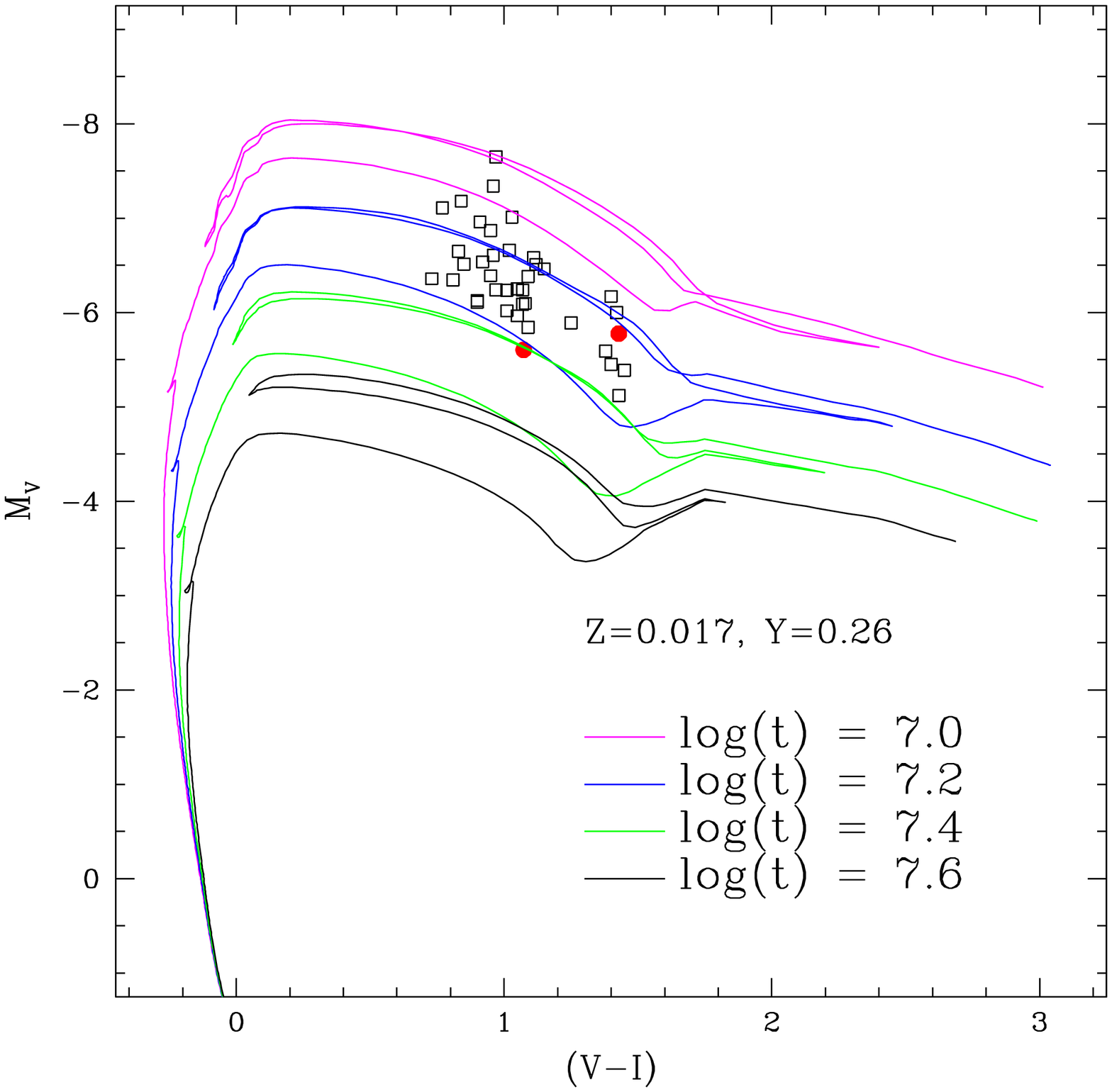} \\ 
  \end{array}$ 
  \caption{Isochrones at four different ages with three metallicity ($Z=0.004$ in the left panel; $Z=0.008$ in the middle panel; and $Z=0.017$ in the right panel) overplotted on CMD with ULPCs taken from \citet[][open squares]{fiorentino2012} and the two M31 ULPCs (red filled circles). These isochrones were taken from \citet{bertelli2009}.} \label{fig_isochrone}
\end{figure*}

Regardless of their potential or problems in future distance scale applications, ULPCs represent a unique probe for studies of stellar pulsation and evolution because they occupy the upper part of the instability strip that has attracted little attention to date. Evolutionary tracks based on stellar evolution models appropriate for ULPCs have been explored in \citet{bird2009} and \citet{fiorentino2012} and will not be repeated here. In Figure \ref{fig_isochrone}, we showed the isochrones adopted from \citet{bertelli2009}\footnote{{\tt http://stev.oapd.inaf.it/YZVAR/}} on the CMD, suggesting that ULPCs are young objects as expected. Nevertheless, detailed investigation of the evolutionary status for ULPCs with theoretical evolutionary tracks and isochrones is beyond the scope of this paper and will be addressed in future work.

\acknowledgments

We thank the referee for valuable input that improved the manuscript. The authors acknowledge the funding from the National Science Council of Taiwan under the contracts NSC101-2112-M-008-017-MY3 and NSC101-2119-M-008-007-MY3. SMK acknowledges the Indo-U.S. Science and Technology Forum for funding partial work carried out in this project. We also thank C. Broeg for sharing his code to calculate the differential photometry, as well as G. Fiorentino, L. Macri and J. Mould for valuable discussions. We acknowledge MIAPP (Munich Institute for Astro- and Particle Physics) for organizing the 2014 ``Extra-Galactic Distance Scale'' Workshop, at which part of this work was conducted.

The Digitized Sky Surveys were produced at the Space Telescope Science Institute under U.S. Government grant NAG W-2166. The images of these surveys are based on photographic data obtained using the Oschin Schmidt Telescope on Palomar Mountain and the UK Schmidt Telescope. The plates were processed into the present compressed digital form with the permission of these institutions.

The National Geographic Society - Palomar Observatory Sky Atlas (POSS-I) was made by the California Institute of Technology with grants from the National Geographic Society.

The Second Palomar Observatory Sky Survey (POSS-II) was made by the California Institute of Technology with funds from the National Science Foundation, the National Geographic Society, the Sloan Foundation, the Samuel Oschin Foundation, and the Eastman Kodak Corporation.

The Oschin Schmidt Telescope is operated by the California Institute of Technology and Palomar Observatory.

The UK Schmidt Telescope was operated by the Royal Observatory Edinburgh, with funding from the UK Science and Engineering Research Council (later the UK Particle Physics and Astronomy Research Council), until 1988 June, and thereafter by the Anglo-Australian Observatory. The blue plates of the southern Sky Atlas and its Equatorial Extension (together known as the SERC-J), as well as the Equatorial Red (ER), and the Second Epoch [red] Survey (SES) were all taken with the UK Schmidt.

Facilities: Palomar Observatory's 1.5 meter Telescope, Lulin Observatory's 1.0 meter Telescope %\facility{PO:1.5m} 

% ===============================================
%               REFERENCE
% ===============================================

\end{document}